\documentclass[]{pasj01}
\draft

\usepackage{lineno}

\begin{document} 


\title{Revealing the dynamics of magnetosphere, atmosphere, and interior of solar system objects with the Square Kilometre Array}

\author{Tomoki \textsc{kimura}\altaffilmark{1}}%
\altaffiltext{1}{Tokyo University of Science, 1-3, Kagurazaka, Shinjuku, Tokyo, 162-8601, Japan}
\email{kimura@rs.tus.ac.jp}

\author{Yuka \textsc{fujii}\altaffilmark{2}}
\altaffiltext{2}{National Astronomical Observatory of Japan, 2-21-1 Osawa, Mitaka, Tokyo, 181-8588, Japan}
\email{yuka.fujii.ebihara@gmail.com}

\author{Hajime \textsc{kita}\altaffilmark{3}}
\altaffiltext{3}{Tohoku Institute of Technology, 35-1, YagiyamaKasumi-cho, Taihaku-ku,Sendai, Miyagi, 982-8577, Japan}
\email{hajimekita@tohtech.ac.jp}

\author{Fuminori \textsc{tsuchiya}\altaffilmark{4}}
\altaffiltext{4}{Tohoku University, Aramaki-aza-Aoba 6-3, Aoba, Sendai, Miyagi, 980-8578, Japan}
\email{fuminori.tsuchiya.a8@tohoku.ac.jp}

\author{Hideo \textsc{sagawa}\altaffilmark{5}}
\altaffiltext{5}{Kyoto Sangyo University, Motoyama, Kamigamo, Kita-ku, Kyoto, 603-8555, Japan}
\email{sagawa@cc.kyoto-su.ac.jp}

\author{the SKA-Japan Planetary Science Team}


\newcommand\YF[1]{\textcolor{red}{[#1]}}
\newcommand\TK[1]{\textcolor{blue}{[#1]}}

\maketitle

\begin{abstract}
Bodies such as planets, moons, and asteroids in our solar system are the brightest objects in the low-frequency radio astronomy at $\lesssim 10 \mathrm{GHz}$. The low-frequency radio emissions from our solar system bodies exhibit varieties of observed characteristics in spectrum, polarization, periodicity, and flux. The observed characteristics are essential probes for explorations of the bodies' magnetosphere, atmosphere, surface, and even interior. Generation and propagation theories of the radio emissions associate the characteristics with fundamental physics embedded in the environments: e.g., auroral electron acceleration, betatron acceleration, and atmospheric momentum transfer. Here we review previous studies on the low-frequency radio emissions from our solar system bodies to unveil some outstanding key questions on the dynamics and evolution of the bodies. To address the key questions by the future observations with the Square Kilometre Array (SKA), we made feasibility studies for detection and imaging of the radio emissions. Possible extensions of the solar system observations with SKA to the exoplanets are also proposed in the summary.

\end{abstract}





\section{Introduction}

Solar system planets and satellites are one of the best-known targets of astronomical observations. 
They were the targets of the first astronomical telescopes, and have now been accessible in a broadest range of wavelengths from X-ray to low-frequency radio wave. 
Astronomical observations have substantially contributed to the study of their detailed properties including atmospheric and magnetospheric structures, in synergy with recent \textit{in-situ} measurements with the spacecraft.
Nevertheless, there remain many mysteries for which new observations are expected to give clues. 
One of such new observations will be enabled by the new low-frequency radio observatory, the Square Kilometre Array (SKA; e.g., \cite{skatechdoc}), with the high sensitivity in frequency bands 0.1-10~GHz. 

The expected scientific goals of SKA for the solar system bodies were broadly reviewed by \citet{butler04}. 
The authors found that SKA observations will have unique contributions to virtually all objects in the solar system, including gas giants, terrestrial planets, moons, and small icy bodies, through non-thermal and/or thermal radio emissions. 
It will be able to characterize synchrotron emission of Jupiter's radiation belt to probe the 3D configurations of energetic particles and magnetic fields, and may also detect the synchrotron radiation from Saturn, Neptune and Uranus for the first time. 
The low frequency with SKA will also probe the deep layers of atmospheres of gas giants and Venus that are optically thick at higher frequencies. 
Combined with SKA's high sensitivity, this allows us to unveil the compositions of deep atmospheres and their horizontal distributions. 
The surface and sub-surface properties of terrestrial planets and relatively major icy bodies can also be studied with the information of spatial distribution. 
Furthermore, SKA can be used to investigate the physical and chemical properties of minor bodies. 

Since then, one has seen substantial advances in our understanding of these planets and satellites. 
For example, recent \textit{in-situ} observations of Jupiter's magnetosphere by the Juno explorer have elucidated the detailed micro-physics for the auroral electron acceleration in the polar region and yielded the detection of localized strong magnetic fields (see \S\ref{sss:juno}). 
Juno also observed deep inside the Jovian atmosphere with collaborations of multi-wavelength observations by space- and ground-based telescopes.
Regarding the terrestrial planets, observations of Venusian atmosphere by the Akatsuki orbiter partially solved a long-standing question about the maintenance mechanism of the very rapid atmospheric circulation (\S\ref{subsub:vensci}). 

The purpose of this paper is to review the specific questions raised from the recent low-frequency radio observations for the solar system bodies and to examine the use of SKA to address some of the extracted questions, quantitatively evaluating the radio detectability with SKA. 
In particular, we focus on investigating the process of particle acceleration in the planetary magnetospheres (\S\ref{s:auroral} and \S\ref{s:synchrotron}), the thermal structure of Venusian lowermost atmosphere (\S\ref{sub:therpla}), and the possible thermal emission from water plumes at the icy moons of the gas giants (\S\ref{ss:icy_body}). Our perspectives for the future SKA observations are summarized in \S\ref{s:summary} with discussions on the impact on the exoplanet in a broader context.

\section{Non-thermal emissions: auroral radio emissions}
\label{s:auroral}

The brightest radio emission in our solar system is the auroral radio emission from the planetary polar region.
In general,  auroras from X-ray to infrared wavelengths (hereafter denoted as``optical auroras'') are emitted from the atmospheric atoms and molecules excited via collision with the energetic electrons precipitating from the magnetosphere. 
A part of the energy in the precipitating electrons is converted to the electromagnetic waves through the plasma instability. That is the auroral radio emissions, which occur at altitudes higher than the optical auroras. 
Because the generation processes of both the polar optical auroras and the aurora radio emissions are similarly related to the precipitating electrons, correlations between those two auroras have long been reported for the most of solar system planets \citep{zarka04}. 
As for Jupiter, $\sim 1$\% of the total energy of precipitating electron is converted to the energy of auroral radio emission \citep{zarka04}. 
The total electromagnetic energy of $\sim$600 TW is extracted from Jupiter \citep{hill01} via the auroral current system driven by the planetary rotation and intrinsic magnetic field energy (Hill current system, see \S \ref{sss:auroral_current_hill}). 200--800 GW out of the total electromagnetic energy is allocated to the optical auroral emissions and only 10--100 GW to the auroral radio emissions \citep{bagenal11,clarke04}. 

\begin{figure}[bt!]
\centering
\includegraphics[width=0.8\hsize]{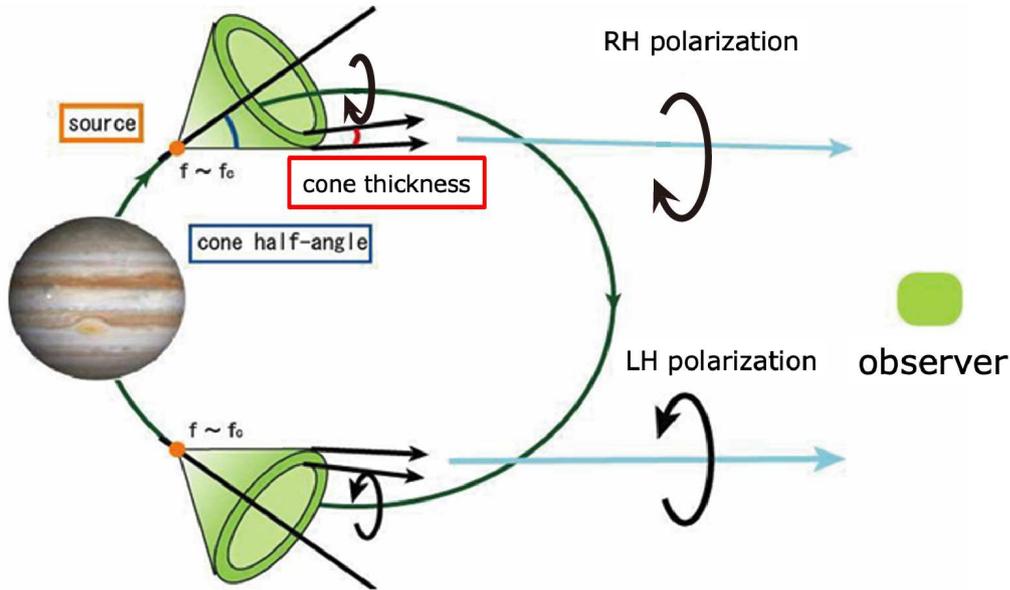}
\caption{A schematic of the thin ``hollow-cone'' shaped beaming of Jupiter's hectometric auroral radio emission HOM, and polarizations with respect to the background magnetic field in the source region and to the observation antenna \citep{itoh08}. \label{fig:itoh_08}}
\end{figure}

From the previous observations and theories, it is highly expected that the source of auroral radio emission is fixed to the planetary magnetic field that connects to the polar optical aurora with an anisotropic beaming angle with respect to the source magnetic field line.
This anisotropy depends on the radio frequency, velocity distribution of source precipitating electrons, and local cyclotron frequency at the radio source region. 
An extreme case for such anisotropy is Jupiter's hectometric auroral radio emission (HOM) that has a thin ``hollow-cone'' shaped beaming with an opening angle of $\sim 10\mathrm{s}^\circ$. 
The radio emissions are refracted and scattered by plasmas in Jupiter's magnetosphere that are supplied from the moons (predominantly Io). 
As a result of the intrinsic anisotropy of emission and the modulation of the ray paths, 
an observer far outside the magnetosphere can observe the auroral radio emission only at $\sim \pm10^\circ$ from the equator (\cite{itoh08}, Figure \ref{fig:itoh_08}). 
On the other hand, Saturn's auroral radio emission (SKR) has a broad ``filled-cone'' shaped beaming of several $10\mathrm{s} ^\circ$, which are observable at almost all latitudes \citep{lamy08}. 
Differences of anisotropy of the hollow- and filled-cones are attributed to the unstable velocity distribution of precipitating auroral electrons (see \S\ref{ss:auroral_generation})

Because the source region of the auroral emission corotates with the planetary magnetosphere, the radio emission observed at a remote observer has a periodic nature. 
Such a periodicity of radio emission can uniquely constrain the rotation period of the interior of gas giants, which cannot be exactly estimated by optical observations due to the dynamics of the overlying atmosphere. 
Indeed, Jupiter's ``true'' rotation period is estimated to be $9^\mathrm{h} 55^\mathrm{m} 29^\mathrm{s}.37$ based on the decametric auroral radio emission (DAM), 
while the rotation period derived from the cloud motion in the equatorial region is to be $9^\mathrm{h} 50^\mathrm{m} 30.0034^\mathrm{s}$ \citep{dessler02}. 


\subsection{Generation process of auroral radio emission}
\label{ss:auroral_generation}


Here we briefly review the generation process of auroral radio emission. 
The generation process theories associate the temporal and spectral features of observed auroral radio emissions with essential characteristics of the source region: e.g., the velocity distribution of precipitating electron and cyclotron frequency of the background magnetic field.

\begin{figure}[htb]
\centering
\includegraphics[width=0.5\hsize]{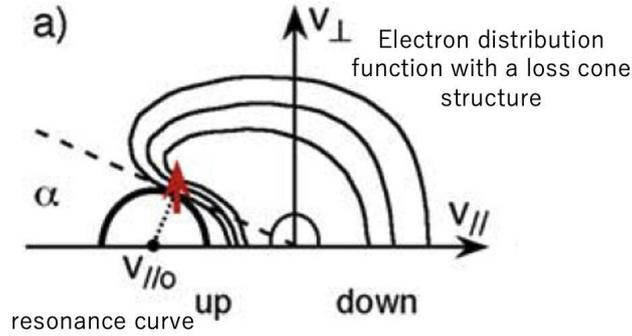}
\caption{
The loss cone structure of electron distribution function and the ``resonance curve'' that is used for estimation of the radio wave growth rate \citep{hess08}. 
Horizontal axis shows the electron velocity parallel to the background magnetic field, and vertical axis shows the perpendicular velocity. 
Thick black line is the resonance curve, and thin black line is the contour of velocity distribution function density with the loss cone structure. 
The phase velocity of circularly polarized and parallelly propagating radio wave that most effectively resonates with the loss cone distribution is referred to as $V_{||0}$,
\label{fig:hess_08}}
\end{figure}

The Cyclotron Maser Instability (CMI) is the excitation process of coherent electromagnetic free space wave (i.e., radio emission) that is the most plausible mechanism of the auroral radio emissions \citep{wulee79}. 
In the CMI process, the gyrating electrons ``resonate'' with the circularly polarized radio wave that oscillates in phase with the gyrating electrons and grows the wave amplitude throughout the energy conversion from the electron to the wave (or which dumps the wave amplitude depending on the resonance condition). 
The ``loss cone'' shaped distributions (Figure \ref{fig:hess_08}) is one of the most effective distribution for the growth of radio wave.
Theoretically estimated beaming and spectrum of the excited radio wave have been found to well explain most of the previous observations \citep{zarka04}. 

The linear growth and dumping rates are estimated by integration of a gradient in the electron velocity phase space density along the ``resonance curve'' (the thick black line in Figure \ref{fig:hess_08}). 
The resonance curve is defined as: 
\begin{equation}\label{eq:resonance}
    \omega - k_{||}v_{||} = \frac{n\omega_c}{\gamma}, 
\end{equation}
where $\omega$ is the wave angular frequency, $k_{||}$ is the wavenumber parallel to the background magnetic field, $v_{||}$ is the parallel electron velocity, $n$ is the resonance order, $\omega_c$ is the local cyclotron angular frequency, and $\gamma$ is the Lorentz factor of electron. The cyclotron angular frequency $\omega_c$ is given with the elementary charge $e$, the background magnetic flux density $B$, and the electron mass $m_e$ as: 
\begin{equation}
    \omega _c = \frac{e B}{m_e} = 28 [ \mathrm{Hz} ] \times \left( \frac{B}{1 [ \mathrm{nT} ] } \right)
\end{equation}

Equation \ref{eq:resonance} corresponds to a quadratic curve that depends on $v_{||}$ and $v_{\perp}$ in the electron velocity phase space.
The growth and dump of radio wave are determined by the positive and negative sign of the gradient integration of distribution function along the resonance curve.
If the integration value is positive, the radio emission is excited via the wave growth, and vice versa. 
The integration of the most stable distribution function, Gaussian, always yields negative, which results in the dump of radio wave. 
The resonance curve depends on the dispersion relation of radio wave (i.e., combination of $(\omega,k)$).
Therefore, radio waves are either excited or dumped with particular combinations of $(\omega,k)$. 
This causes a variety of properties of the spectral structures and beaming with respect to the background magnetic fields.
These properties can be theoretically estimated based on the numerical integration of distribution function assuming the resonance curve. 

The CMI process requires that in the radio source region the source electrons are collisionless and gyrate at the cyclotron frequency $\omega_c$ much greater than the electron plasma frequency $\omega_p$ (i.e., $\omega_p/\omega_c \ll 1$), where
\begin{equation}
\omega _p = 2 \pi f_p, \;\;\;\; f_p = \sqrt{\frac{n_e e^2}{\pi m_e}} = 8.979~ [ \mathrm{kHz} ] \times \left( \frac{n_e}{1~ [ \mathrm{cm}^3 ] } \right)^{1/2} \label{eq:nu_plasma}
\end{equation}
with the electron number density $n_e$.
This requirements corresponds to the condition in which the electron gyrates much faster than the plasma oscillation and effectively resonates with the radio wave. 
This condition is reasonably satisfied above the polar region of a strongly magnetized planet. 

The resonance condition in Equation \ref{eq:resonance} can be reduced to $\omega \sim \omega _c $ in the case of non-relativistic electrons and the resonance order $n=1$.
This means the excited radio wave frequency almost equals the cyclotron frequency. 
With $\omega \sim \omega_c$, the magnetic flux density in the radio source region can be estimated from the remotely observed radio frequency: 
\begin{equation}\label{eq:cyclo}
    B \sim \frac{m_e\omega}{e} 
\end{equation}
Note that this approximation is not valid for some of radio emission that are excited via CMI by the relativistic ($>$MeV) electrons, where the Lorentz factor is much greater than unity (\cite{kimura11b}).
In the relativistic case, the excited radio emission frequency is significantly smaller than the local cyclotron frequency with the dependence on the electron energy.

\subsection{Energy source of auroral radio emission}
\label{ss:auroral_energy}

The precipitating electron that is the energy source of auroral radio emission dynamically changes in the flux and energy depending on the magnetospheric dynamics.
The intensity and spectrum of excited auroral radio emission respond to the changes in the precipitating electron flux and energy.
In this section, three energy sources for the auroral electron acceleration are introduced briefly.

\subsubsection{Hill current system}
\label{sss:auroral_current_hill}

The most effective acceleration process for the auroral electron is the field aligned electric potential associated with the current system that connects the ionosphere and magnetosphere. 

\begin{figure}[tb]
        \begin{center}
        \includegraphics[width=0.5\hsize]{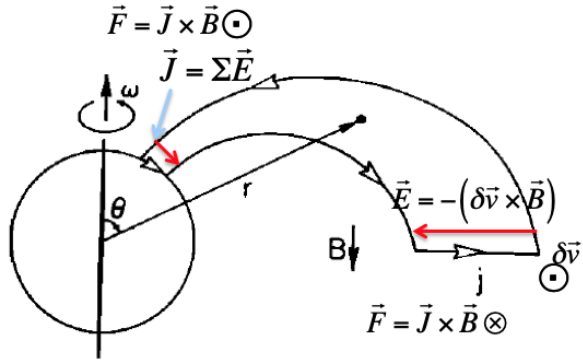}
        \caption{Schematic of the Hill current system \citep{hill79}. White head arrow and red arrow indicate the magnetosphere-ionosphere coupling current and dynamo electric field in the magnetosphere, respectively. \label{fig:hill_79}}
        \end{center}
\end{figure}

At rapidly rotating magnetized gas giants like Jupiter and Saturn, the Hill current system is created through the dynamo process driven in their rotating magnetospheres. 
This current system requires (1) a rapid planetary rotation, (2) a strong intrinsic magnetic field, and (3) a magnetosphere filled with the space plasma: e.g., at Jupiter, the rotation period of $\sim 10$ hr, the surface magnetic field of $\sim 5$ G around the equator, and the plasma density of $2,000 ~\mathrm{cm^{-3}}$ at 6 $\mathrm{R_j}$ supplied from volcanoes at Io, respectively. 
The plasmas supplied from Io sub-corotate because of angular momentum conservation as they are centrifugally transferred outward.
An azimuth velocity of corotation lag $ \delta \mathbf{v} $ creates the ``dynamo electric field'' $\mathbf{E}=-\delta \mathbf{v} \times \mathbf{B}$ as in Figure \ref{fig:hill_79}, where $\mathbf{B}$ is the magnetic flux density at the magnetosphere.
The dynamo electric field is imposed on the polar ionosphere with a finite electric conductivity, which generates the electric current circuit in the meridian plane that couples the ionosphere with the magnetosphere. 
The Lorentz force ($\mathbf{J} \times \mathbf{B}$) created by the current system extracts the planetary rotation angular momentum and transfers it to the magnetosphere. 

The Hill current reaches up to 10s MA in total at Jupiter (\cite{hill01}). 
The field-aligned electric current is required to maintain its flux (i.e., current density) in the polar magnetosphere above the ionosphere, which leads to the field-aligned electric field, e.g. so-called the inverted-V field structure (e.g., \cite{cowley01}). 
The electron acceleration energy by the field-aligned electric field reaches up to 100s keV or possibly up to the MeV range and precipitates the electrons into the atmosphere (e.g., \cite{mauk17}). 
The optical and radio auroral emissions are excited by the precipitating electrons.
The total energy consumed for the radio emission excited by the field-aligned acceleration of the Hill current system is estimated to be 100 GW (\cite{zarka07}).

\subsubsection{Io-Jupiter current system}
\label{sss:auroral_current_io}

At the Jupiter system, the corotating magnetic field moves through Io's volcanic atmosphere at a relative azimuth velocity of $|\mathbf{v}|\sim54$ km~s$^{-1}$, which creates the dynamo electric field of $\mathbf{E} = - \mathbf{v} \times \mathbf{B}$ that generates the current in Io's ionized atmosphere, which couples Io with Jupiter's polar region \citep{thomas04}.
Directions of the dynamo electric field and current at Io are radially outward. 
The Io-Jupiter current system has the total current of 1 MA and excites the auroral radio emission in the same manner with the Hill current system. 
The total energy consumed as the auroral radio emission in the Io-Jupiter current system is estimated to be a few GW. 
The auroral radio emission associated with the Io-Jupiter current system, known as Io-DAM, is excited above the polar footpoint of the magnetic field line that threads Io.
Occurrence of Io-DAM therefore depends on the orbital phase of Io as well as on Jupiter's rotation phase (e.g., \cite{dessler02}).

\subsubsection{Solar wind}
\label{sss:auroral_current_sw}

\begin{figure}[htb]
\centering
\includegraphics[width=0.7\hsize]{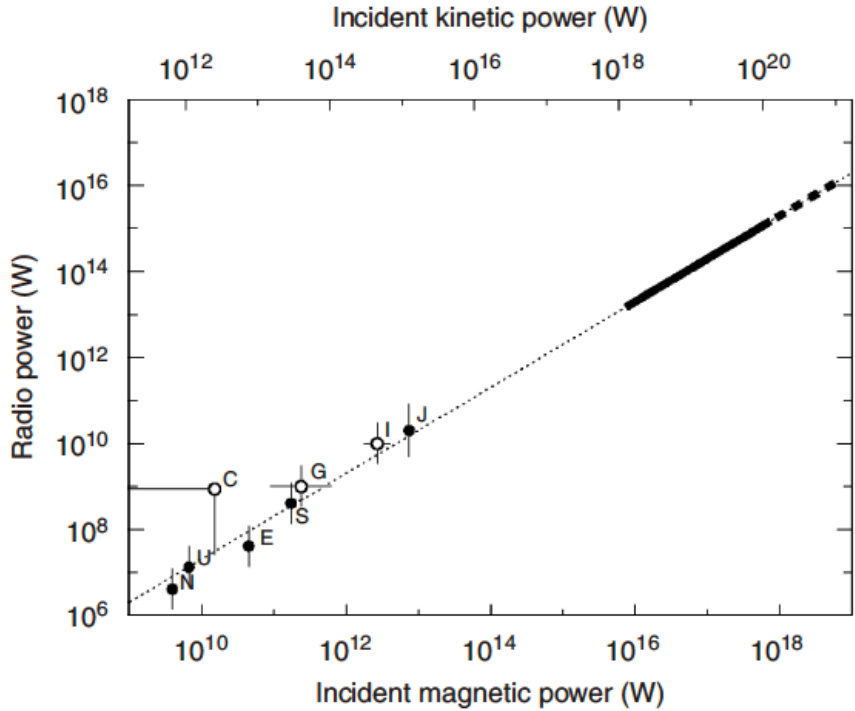}
\caption{Scaling law of the planetary auroral radio flux \citep{zarka07}. This diagram shows the relation of auroral radio fluxes with the  energies input to the planetary magnetospheres. Horizontal axis shows the electromagnetic or kinetic energy of the solar wind, and vertical axis shows the flux density emitted from the planets and moons in our solar system. N,U,C,E,S,G,I, and J denote, Neptune, Uranus, Callisto, Earth, Saturn, Ganymede, Io, and Jupiter, respectively.} \label{fig:zarka_07}
\end{figure}

The solar wind, which is the plasma flow originating from Sun is also a significant energy source for the auroral radio emission and other electromagnetic dynamics in the magnetosphere, ionosphere, and atmosphere. 
A small fraction (typically $\sim 1 \%$) of the total electromagnetic energy deposited in the solar wind is input to the magnetosphere and drives the convection of plasma, which ultimately creates the current system in the magnetosphere. 
For example, at Jupiter's magnetosphere, the solar wind energy input is estimated to be 130 TW \citep{bagenal11}, while the total energy extracted from Jupiter's rotation and intrinsic magnetic field is estimated to be $\sim 600$ TW stored in the magnetosphere \citep{hill01}.
In the similar manner to the Hill current system, the solar wind energy creates the current system that accelerates the auroral electron above the polar region. 
The input rate of solar wind energy depends on the Poynting flux of the solar wind and a cross section of the magnetosphere. 
\citet{zarka07} found a scaling law between the energy input from the solar wind and the flux density of the auroral radio emission (Figure \ref{fig:zarka_07}). 
This is the ``Bode's law for the planetary radio emissions'', which is recently used for the estimation of auroral radio flux of the exoplanets. 
The auroral radio flux varies in response to the solar wind variability.
For example, \citet{gurnett02} and \citet{prange04} observed the auroral radio flux of Jupiter intensified by a factor of 3--5 from the quiet level when a solar wind shock structure compressed at Jupiter's magnetosphere.

\subsection{Recent insights into energy source of auroral radio emission}

The energy sources of auroral radio emission reviewed in the previous sections dynamically change in time and space, depending on the environments of magnetosphere.
Here we briefly review the recent studies on the dynamics of the energy source of auroral radio emission.

\subsubsection{Mass transfer in magnetosphere}
\label{sss:mass_transfer}

The auroral current systems reviewed in \S\ref{ss:auroral_energy} are generated by the circulation of plasmas in the magnetosphere. 
Mass distribution of the circulating plasmas is always controlled by the balance between the source and loss processes. 
For example at Jupiter, the plasma mass supplied from Io's volcanoes possibly experiences the centrifugally-driven interchange motion in the plasma torus, followed by the radial transport to the middle and outer magnetosphere \citep{Horne2008}.
The total mass of magnetosphere distributed by the radial transfer is estimated to be $1.5 \times 10^{9}~\mathrm{kg}$.
In the quasi-steady state, the total mass is maintained by the balance of mass source process from Io's volcanic gas at a few 100s--1000 kg~s$^{-1}$ with the mass loss process by the plasma ejection from the magnetosphere to the interplanetary space (e.g., \cite{bagenal11}).
The mass balance and associated plasma circulation in the magnetosphere temporally changes in response to the solar wind energy input and the eruption rate of water and volcanic gas from the moons \citep{kimura18}. 
This finally leads to change in the auroral current system and resultant electron acceleration. 
Actually, the transient UV auroral burst at a period of $\sim$ 2--3 days associated with the plasmoid eruption got intensified when Io's volcano indicated a large eruption signal with a mass loading rate enhancement by a factor of 1.5 from the steady state \citep{kimura18}.
The UV auroral burst follows the transient auroral radio emission with the energetic electron injections possibly associated with the magnetotail reconnection (e.g., \cite{louarn14}).

\subsubsection{Solar EUV dependence}
\label{sss:solar_euv}

\citet{kimura13} found the seasonal variation in Saturn's auroral radio flux that is likely caused by the significant seasonal change in the solar UV flux irradiated to Saturn's atmosphere due to a large axial tilt of 26.7$^\circ$. 
The auroral radio flux emitted from the summer hemisphere is 100 times greater than that from the winter hemisphere. 
Based on the spectral structure of auroral radio emission, \citet{morioka12} suggested the vertical extent of Earth's auroral acceleration region clearly dependent on the seasons due to a large axial tilt of 23.4$^\circ$.
Although the cause of seasonal variations in the flux density and spectrum is still unknown, the solar UV flux irradiated to the ionosphere likely changes in correlation with the planetary orbital motion that modifies the ionospheric conductivity, which leads to modification of the auroral current system and radio emission. 
The previous studies interpreted that the seasonal variability in the flow patterns of ionospheric plasma and neutral atmospheric wind may also change the auroral current system and resultant auroral radio emission (e.g., \cite{jia12}).

\subsubsection{Results from Juno: essential physics of auroral electron acceleration}
\label{sss:juno}

\begin{figure}[htb]
        \begin{center}
        \includegraphics[width=0.8\hsize]{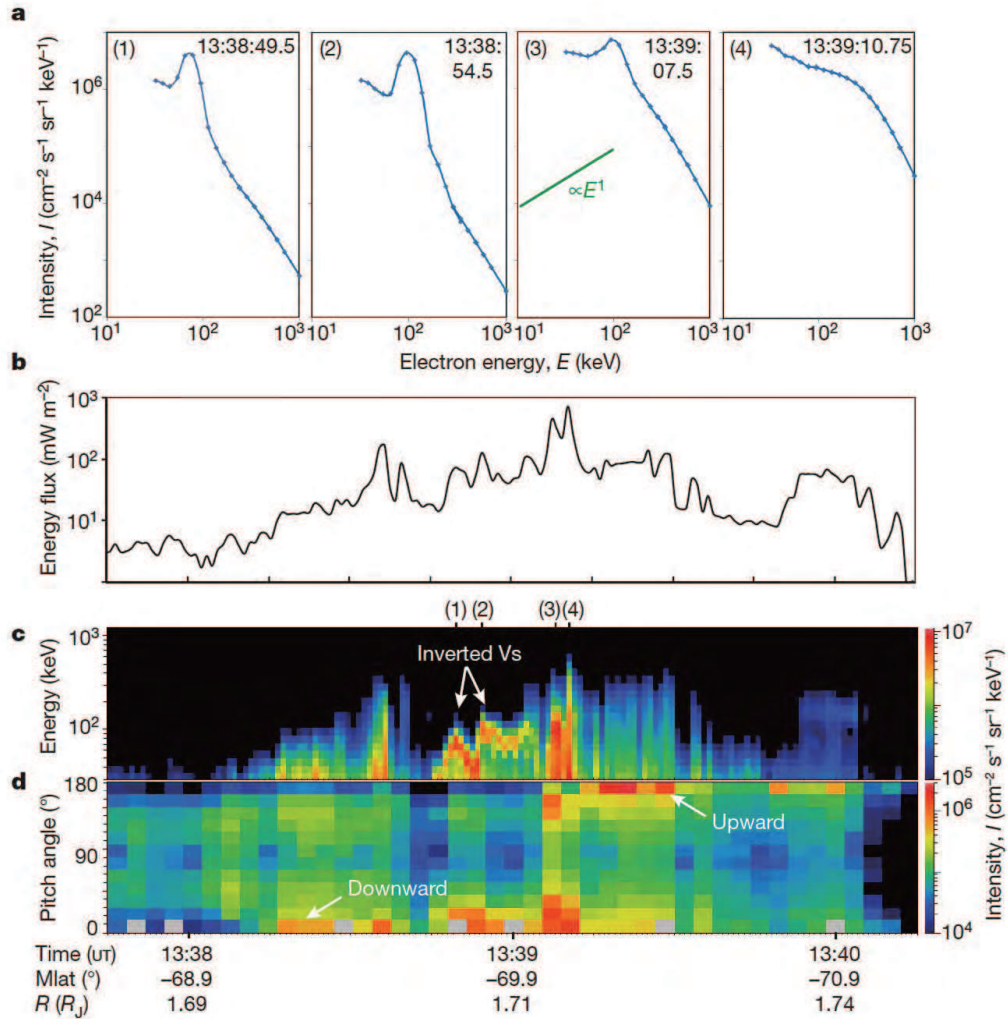}
        \caption{Energy spectrum, energy flux, and energy-time diagram for the energetic auroral electrons measured with the energetic particle detector onboard the Juno explorer above the main emission region \citep{mauk17}. 
        (a) Differential energy flux spectra at periods labeled (1), (2), (3), and (4) in panel (c). (b) Downward precipitating electron energy flux. (c) Energy-time spectrum of downward precipitating electron. (d) Pitch angle distribution of the electrons at 10-1,000 keV.
        \label{fig:mauk_04}}
        \end{center}
\end{figure}

Recently, the Juno explorer made surprising discoveries on the micro-physics for the auroral electron acceleration in Jupiter's polar region based on its \textit{in-situ} measurements.
There have been strongly expected that the inverted-V field aligned electric potential structure is the dominant auroral acceleration process at Jupiter as well as Earth (e.g., \cite{cowley01}).
However, the Juno \textit{in-situ} observation above the main emission region clearly showed that the inverted-V structure was a spatially and energetically minor component (Figure \ref{fig:mauk_04}, \cite{mauk17, mauk20}).
In \citet{mauk17} and \citet{mauk20}, the most prominent process was found to be the stochastic acceleration that is likely driven by the dispersive Alfv\'{e}n wave  \citep{saur18} propagating from the equatorial magnetosphere to the polar auroral region. 
The energy flux of the precipitating auroral electron is estimated from the Juno \textit{in-situ} data to be $\sim 10^{-1} ~\mathrm{W~m^{-2}}$ for the inverted-V structure, while that to be $\sim 1~\mathrm{W~m^{-2}}$ for the stochastic acceleration region (Figure \ref{fig:mauk_04}b, \cite{mauk17}). 

Some pioneering studies \citep{saur03, tao15} already pointed out that the Alfv\'{e}n waves are excited from the MHD turbulence excited in the equatorial middle magnetosphere at 20--30 $\mathrm{R_j}$ and could work as a significant energy source of electrons after propagation from the equatorial middle magnetosphere to the polar auroral region. The polar stochastic acceleration newly discovered by Juno would be a part of this process. The relationships of the Alfv\'{e}nic stochastic acceleration with the traditional Hill current system (\S\ref{sss:auroral_current_hill}) and Io-Jupiter current system (\S \ref{sss:auroral_current_io}) are still unknown. However, the magnetic helicity of Alfv\'{e}n wave is equivalent to the field-aligned current in the MHD theory, which is suggestive of the global auroral current systems involving the Alfv\'{e}n wave in microscopic view. \citet{saur03} estimated the total Poynting flux carried by the magnetospheric turbulence toward the auroral region to be 30 TW at 18--30 $\mathrm{R_j}$, which is roughly a half of the total energy carried by the Hill current system from the equatorial middle magnetosphere at 21 $\mathrm{R_j}$. This implies that the Alfv\'{e}n wave carries a significant amount of the electromagnetic energy as the electric current from the magnetosphere to the auroral region. The Alfv\'{e}nic acceleration is now being numerically solved along the field line between the magnetosphere and ionosphere involving the kinetic effect of energetic electrons interacting with the dispersive Alfv\'{e}n wave \citep{damiano19}. 

Source process of the MHD turbulence in the equatorial magnetosphere is also the big unsolved problem. The study for the  Alfv\'{e}nic acceleration based on the wave-particle interaction theory \citep{saur18} supposes the interchange motion (\S\ref{sss:mass_transfer}) in the equatorial magnetosphere for the driver of the Alfv\'{e}n wave propagating to the polar auroral region. However, the excitation process of Alfv\'{e}n wave from the MHD turbulence has not benn demonstrated yet for Jupiter's centrifugally-driven system.

The micro-physics of excitation process for the auroral radio emissions are quantitatively investigated based on the combination of the wave generation theory with the Juno \textit{in-situ} energetic electron measurement. \citet{louarn17} and \citet{louarn18} theoretically demonstrated HOM to be excited with a sufficient wave growth rate from the observed electron velocity distribution with the conic and loss cone structures, which are consistent with the Alfv\'{e}nic stochastic acceleration process. 
\citet{louarn17} obtained a typical e-folding times of $10^{-4}$ s for the wave amplitude growth, which leads to an amplification path of $\sim$1000 km, likely confined within the radio source region.
\citet{louis21} comprehensively investigated the latitudinal distribution of each component of the auroral radio emissions, which will constrain the location, beaming, and electron distribution function of the auroral radio source.

In addition to the main emission region, Juno also found that the mono-directional upward MeV electron beam and downward MeV ions in the polar cap region \citep{mauk18, mauk20}. 
These energetic particle beams are likely associated with the quasi-periodic auroral radio bursts and X-ray auroras excited in the polar cap region with periods of a few to 10s minutes \citep{yao21}.

\subsection{Imaging of Jovian auroral radio emission source with SKA}
\label{ss:detectability}

Some key questions are extracted from the reviews for the auroral acceleration processes described in the previous sections:
\textbf{
\begin{itemize}
    \item How are the Alfv\'{e}n wave and associated MHD turbulence excited by the interchange motion in the equatorial magnetosphere? (\S \ref{sss:mass_transfer} and \ref{sss:juno})
    \item How do the Alfv\'{e}nic and inverted-V accelerations change in time and space at the ionosphere and magnetosphere? (\S\ref{sss:juno})
    \item How do the energy sources control the auroral accelerations? (\S\ref{sss:auroral_current_hill}, \ref{sss:auroral_current_io},  \ref{sss:auroral_current_sw}, \ref{sss:mass_transfer}, and \ref{sss:solar_euv})
\end{itemize}
}
Here we propose future observations with SKA to address these key questions.

As proposed by \citet{morioka12}, the spatial extent of auroral radio emission in the source region directly corresponds to that of the precipitating auroral electrons.
Imaging of the auroral radio emission in the source region is essential to unveil the auroral acceleration processes such as the Alfv\'{e}nic and inverted-V accelerations and associated magnetospheric dynamics.
For example, at Jupiter, the maximum magnetic flux densities in the polar auroral region have been estimated to be less than $\sim 10$ G from the previous observations (e.g., \cite{clarke04}), which corresponds to the local cyclotron frequencies less than $\sim 30$ MHz. 
This means that the main structure in the auroral radio spectrum should be below $\sim 30$ MHz, which is lower the observable frequency of SKA.
However, there is a strongly magnetized region called the ``magnetic anomaly'' with a surface magnetic flux density more than 20 G recently found by the \textit{in-situ} measurements with Juno \citep{connerney18}. 
The auroral radio emission excited just above the magnetic anomaly would have frequencies greater than 50 MHz. 
This maximum frequency is the highest value compared to those have been observed by the previous studies (e.g., \cite{zarka04}).
The flux density at 50 MHz would be significantly lower than the main component of radio spectrum below 50 MHz.
The spectral observation with the world-highest sensitivity at $> 50$ MHz with SKA-Low will potentially make the imaging of Jupiter's auroral radio source as quantitatively discussed later. 
The flux density distribution along the source field line tells us the vertical profile of energetic auroral electrons, field-aligned electric field, and their dynamics.
The vertical profile measurements is not feasible for remote sensing at other wavelengths because of no emissions directly from the acceleration region.
Based on such a vertical profile, we can discuss the source process of field-aligned electric field, the magnetosphere-ionosphere coupling process, and their temporal activities. 
In particular, the Alfv\'{e}nic acceleration region found with Juno (\S\ref{sss:juno}) is expected to have the wavy structure along the auroral field line, which could be imaged by the high spatial resolution imaging with SKA.

\subsubsection{Detectability: SKA1 and 2}

Based on the Juno \textit{in-situ} measurements by \citet{connerney18}, the magnetic footpoint of Io at Jupiter's polar ionosphere is found to be threading the magnetic anomaly. 
The Io's magnetic footpoint above the  anomaly with $>$ 20 G at surface would emit the DAM radio emission above 50 MHz, which is associated with the Io-Jupiter current system. 
The DAM radio flux spectrum is largely dependent on the frequency above 50 MHz, which ranges from $1$ Jy to $10^6$ Jy \citep{zarka04}.
The spatial extent of magnetic anomaly on the surface likely has a diameter of $\sim 1 R_\mathrm{j}$, which corresponds to 20 arcsec as seen from Earth around Jupiter's opposition. 
The vertical extent of auroral radio source is unknown, which here we assume the same spatial scale as the surface magnetic anomaly.
With the 1hr-exposure time at 60 MHz and the beam FWHM of 23.5 arcsec, the detection limit of SKA1-Low is 163$\mu$Jy/beam \citep{skatechdoc}, with which we can monitor the auroral radio source by second-by-second snapshots. 
Note that the spatial extent of radio source is comparable to the SKA beam size. 
We could resolve the extended radio source region larger than the beam size during activation of the field aligned electric field. 
The corotation motion of radio source is resolved enough with the current SKA beam size.
Rapid temporal timescales around 10s sec in the radio flux will be detected with the enough signal-to-noise (S/N) ratio estimated above.
This high temporal resolution corresponds to the shortest timescales of macroscopic magnetospheric activities that have ever been detected: e.g., the periodicity of Alfv\'{e}n wave propagating in the magnetosphere. 
The SKA1-Low imaging will detect the most rapid magnetospheric activities, which contributes to understanding the field-aligned electron acceleration process generated by the Alfv\'{e}n wave and other essential physics (see \S\ref{sss:juno}).
The SKA2 with the improved spatial resolution 10 times better than SKA1 will split the radio source region into 10 pixel $\times$ 10 pixel, which resolves the structure of field-aligned electric field. 
By monitoring the temporal variability in the spatial extension and its intensity of auroral radio source region from 10 hours to a month, we can investigate the energy source for the Alfv\'{e}nic acceleration, which could be associated with the Hill current system (\S\ref{sss:auroral_current_hill}), Io-Jupiter current system (\S\ref{sss:auroral_current_io}), solar wind (\S\ref{sss:auroral_current_sw}), mass transfer in the magnetosphere (\S\ref{sss:mass_transfer}), and solar EUV (\S\ref{sss:solar_euv}).

\section{Non-thermal emissions: synchrotron radiation}
\label{s:synchrotron}

\subsection{Planetary radiation belt and acceleration processes}
\label{ss:radiationbelt}
\begin{figure}[b!]
\centering
\includegraphics[width=0.8\hsize]{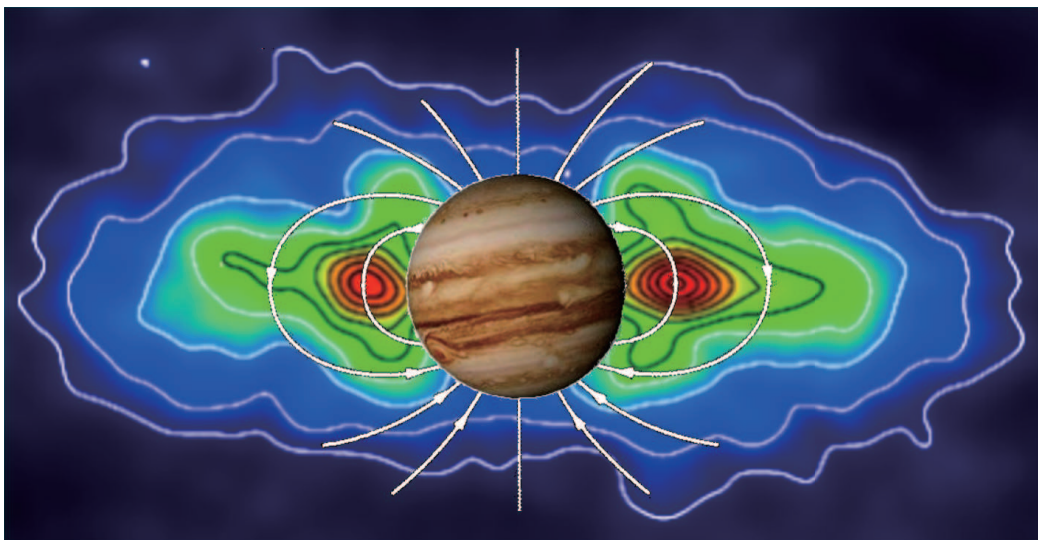}
\caption{Spatial distribution of Jupiter's synchrotron radiation at 610 MHz obsesrved by GMRT. \label{fig:jsr}}
\end{figure}

The radiation belt commonly exists around the magnetized planets. Jupiter has the most intense radiation belt in the solar system, where the electron energy reaches 50-100 MeV \citep{bolton2002}. 
These relativistic electrons move along a spiral path in a magnetic field line under the action of Lorentz force and generates synchrotron radiation. In the case of Jupiter, the synchrotron radiation is observed by a ground-based radio telescope in a frequency range from a few tens of MHz to a few tens of GHz. 
Figure \ref{fig:jsr} shows a spatial distribution of the synchrotron radiation observed with the Giant Metrewave Radio Telescope (GMRT). This map clearly illustrates the concentration of electrons around the magnetic equator, which is indicative of the dipolar nature of the magnetic field. The spatial extent of the synchrotron radiation is limited to a radial distance around 3 $R_{j}$ for the current sensitivity of the radio telescope. 
The core region of this emission region is located around 1.5 $R_{j}$ from Jupiter, while we can also see the high latitude components. 

Synchrotron radiation intensity $P$ [W/Hz] at $\nu$[MHz] emitted from an electron with the energy $E$ [MeV] under the magnetic field strength $B$ [G] is expressed as, 
\begin{equation}\label{eq:sync}
   P \simeq 2.34\times 10^{-29} BF(\nu/\nu_c) \sin\theta \,\, \mathrm{[W/Hz]}
\end{equation}
where, 
\begin{eqnarray}
  F(x)=x\int_0^{\infty} K_{5/3}(\eta)d\eta, \\
  \nu_c=16.08 B E^2 \sin\theta,
\end{eqnarray}
and $K_{5/3}$ is the modified Bessel function. The $\theta$ represents the angle between the electron velocity vector and the magnetic field, which is called the pitch angle. 
When the distance between Jupiter and Earth is 4 AU, the total intensity from Jupiter's synchrotron radiation observed at Earth is about 5 Jy at 1.5 GHz. 
As shown in equation (\ref{eq:sync}), the flux variation of Jupiter's synchrotron radiation is caused by changes in the electron energy, pitch angle, and the spatial distribution of relativistic electrons in the magnetosphere. Compared to other electromagnetic radiations, Jupiter's synchrotron radiation is the effective probe of the spatial distribution and the dynamics of Jupiter's radiation belt.

The transport of electrons in the magnetosphere is approximated by the radial diffusion process. During their inward diffusion, perpendicular energy of the electrons increases because of the conservation of the first adiabatic invariant ($W_\perp/B$, so-called a magnetic moment, where $W_\perp$ is the electron energy perpendicular to the magnetic field). This ``heating process'' is also known as the betatron acceleration. The pitch angle distribution of the radiation belt electron is confined to around 90 degrees because of the betatron electron acceleration perpendicular to the background magnetic field. Jupiter's synchrotron radiation is therefore emitted in the direction perpendicular to the field line with the largest brightness for the edge-on observation. If the magnetic axis is not aligned to the rotational axis, modulation of the synchrotron radiation appears. Jupiter's magnetic axis is inclined by about 10 degrees and apparent intensity modulation of the synchrotron radiation is about 10\%, which is called the beaming curve. The amplitude of the beaming curve depends on the pitch angle distribution of the relativistic electrons. The more confined to 90 degrees the pitch angle distribution is, the larger the amplitude of the modulation is \citep{miyoshi1999}. These physical characteristics of Jupiter's synchrotron radiation were utilized to measure the rotation period, the inclination angle of the magnetic axis, and the magnetic field strength of Jupiter's magnetosphere.

The sources of the radiation belt electrons are the solar wind, the planetary atmosphere, and the gas erupted from the satellites. The electrons experience some acceleration processes and obtain relativistic energy during the radial diffusion \citep{lejosne2020}. In Earth's magnetosphere, radial diffusion is driven by fluctuating Ultra-Low Frequency (ULF) wave and/or substorm induced convection electric fields. However, such processes are ineffective in Jupiter's magnetosphere, because of the large scale of the Jovian magnetosphere, where the ULF waves and convection electric field are not spatially confined with respect to the relativistic particles slowly drifting in the azimuthal direction. 
The radial diffusion in the inner Jovian magnetosphere is assumed to be driven by ionospheric dynamo wind
\citep{brice1973, goertz1979, depater1990}. 
Therefore, the solar UV heating of the ionosphere is one of the important control factors for Jupiter's radiation belt. This scenario is supported by the observations of the time variation and spatial distribution of the synchrotron radiation \citep{miyoshi1999,tsuchiya2011,kita2013,kita2015}.
Outside of Io's orbit, electric field fluctuations driven by the interchange instability \citep{southwood1987,kollmann2018} and by the dawn-dusk electric field \citep{murakami2016, han2018} are the driver for the radial transport and adiabatic heating of electrons. The dawn-dusk electric field is modulated by the solar wind dynamic pressure \citep{murakami2016}, and the year order variation of the synchrotron radiation can be explained by the change of radial diffusion coefficient driven by the dawn-dusk electric field \citep{han2018}. 

Although the betatron acceleration is a major acceleration mechanism in Jupiter's radiation belt, it is not enough to achieve the observed particle energy. In addition to that, all intermediate (weeks-months order) and long (year order) time scale variation cannot explained by the solar UV/EUV heating and the solar wind. Other candidates for the additional acceleration process are pre-heating in the outer magnetosphere and the wave-particle interaction in the middle magnetosphere.
The recent Juno spacecraft observations show that there exist persistent upward energetic electron beam at the Jupiter polar region 
\citep{mauk17,ebert17,clark17}. 
As the polar region is mapped to the outer magnetosphere through the magnetic field line, these electrons will be a source of energetic electrons in the outer magnetosphere. When these electrons move inward, they are further accelerated through the betarton acceleration process and their energy eventually become relativistic in the radiation belt (see also \S\ref{sss:juno}).
It is also expected that the wave-particle interaction also plays an important role in the acceleration process of the radiation belt electrons. Whistler mode wave-particle interaction is an important acceleration process in the radiation belt of the Earth \citep{miyoshi2018}. Whistler-mode chorus emission is one of the plasma waves, which has a fine spectral structure. In the case of Jupiter, Voyager and Galileo identified similar plasma waves like those exist in the Earth. The statistical analysis of the plasma wave observed by Galileo shows that the whistler-mode chorus is enhanced between the orbit of Ganymede and Io, and its peak is located around Europa's orbit \citep{katoh2011}. It has not been revealed that how such a plasma wave activity changes the radiation belt in the intermediate (weeks-months order) and long (year order) time scale. The whistler-mode chorus is excited by the hot electron injected into the inner magnetosphere \citep{Horne2008}, which is related to the outward transport of the heavy plasma ejected from Io by the interchange instability \citep{Yoshioka2014}. It is still an ongoing study that how volcanic activity at Io affects the acceleration process of the radiation belt. If we can observe the synchrotron radiation between the orbits of Io and Europa, it will give us a clue to understanding the electron acceleration mechanism in the strongest radiation belt in the solar system.
\subsection{Observation of planetary synchrotron radiation}
\label{ss:obs_jsr}
\begin{figure}[b!]
\centering
\includegraphics[width=1.0 \hsize]{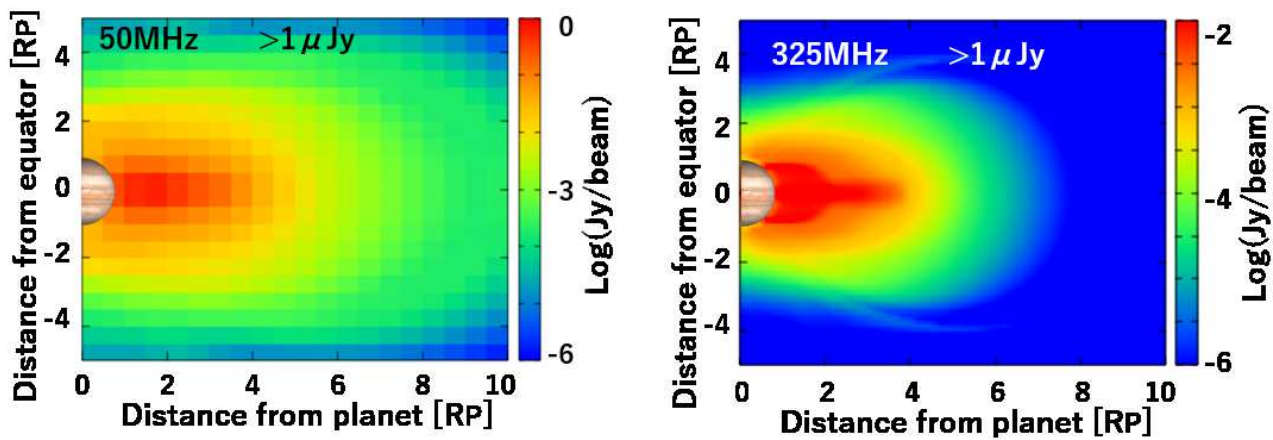}
\caption{Spatial distribution of Jupiter's synchrotron radiation at 50 MHz and 325 MHz with dipole magnetic field and particle distribution model by \citet{garrett2005}.
}\label{fig:ska_jsr}
\end{figure}

Here are the key questions which will be addressed with SKA observations.
\textbf{
\begin{itemize}
    \item How does the pre-heating in the outer magnetosphere control Jupiter's radiation belt?
    \item What is the main source of electric field perturbation for the betatron acceleration through the radial diffusion in the radiation belt.
    \item How does the wave-particle interaction change the radiation belt change the radiation belt.
\end{itemize}
}

The unprecedented high sensitivity of SKA will allow us to probe the weak synchrotron radiation of Jupiter that has not been detected yet. In other words, we can study the radiation belt electrons in a broader region. 
Figure \ref{fig:ska_jsr} shows Jupiter's synchrotron radiation at 50 MHz and 325 MHz. Io and Europa are located at 5.9 $R_{j}$ and 9.4 $R_{j}$, respectively, where the apparent diameter of Jupiter is 30-50 arcsec. Spatial resolutions of SKA 1-low at 50 MHz and 300 MHz are 23.5 arcsec and 4.6 arcsec, respectively  \citep{skatechdoc}. The pixel size of Figure \ref{fig:ska_jsr} is almost the same as the expected beam size. Based on the previous observations of Jupiter's synchrotron radiation, Jupiter's radiation belt varies on timescales of days, and therefore, the integration time of 1-hour is reasonable to detect the time variation. Minimum detectable radio flux for 1-hour integration at these frequencies are 163 $\mu$Jy and 11 $\mu$Jy, respectively \citep{skatechdoc}. Comparing these sensitivities with Figure \ref{fig:ska_jsr}, SKA1 low can detect sychrotron radiatino up to Io orbit at 325 MHz and up to Europa orbit at 50 MHz. 

Here, we show the idea on how to reveal the physical characteristics of the three acceleration processes, pre-heating, betatron acceleration, and wave-particle interaction with SKA, and give us the comprehensive view of the plasma-magnetosphere interaction that produces the radiation belt. 

SKA observations will work in synergy with the in-situ observations by JUICE (to be launched in 2023) and Europa Clipper (to be launched in 2024). While SKA observations can monitor the large-scale structure of the radiation belt, it cannot observe the outer magnetosphere where the pre-heating processes occurs. On the other hand, in-situ observations can observe particle energy and distribution where the pre-heating process occurs, while they cannot obtain the global properties of the radiation belt. The study can be further complemented by the observations of the time variability of the aurora activity by observations of infrared/ultraviolet/radio emissions. Combining these measurements with the synchrotron radiation observed with SKA, we can discuss how they are related to each other. 

It is obvious that the betatron acceleration plays an important role in Jupiter's magnetosphere, however, the source of the electric field perturbation, which causes the radial diffusion in the radiation belt, is still unclear. The electric field perturbation in Jupiter's upper atmosphere is thought to be a major source for radial diffusion, and it is strongly related to the solar UV/EUV. In addition to that, the dawn-dusk electric field is also an important source for the radial diffusion \citep{han2018}, which is related to the solar wind dynamic pressure. The radial diffusion coefficient for the perturbation in Jupiter's atmosphere is proportional to $L^{3}$, where $L$ is the L-value of intrinsic dipolar magnetic field of Jupiter, whereas that of the dawn-dusk electric field is proportional to  $L^{6}$. SKA can observe a wide radial range of synchrotron radiation, which is the essential information to discuss the source of the electric field perturbation based on the spatial change in the synchrotron radiation radio flux and the correlation with the solar wind and solar UV/EUV. 

Although the whistler-mode wave was already identified in Jupiter's middle magnetosphere, it is not understood whether the wave-particle acceleration is an effective acceleration process for Jupiter's radiation belt. SKA can see the wave-particle acceleration region, around Europa orbit, so we can evaluate the time variation of the high-energy electron around the middle magnetosphere which is likely associated with the wave-particle acceleration. Simultaneous observation with JUICE and Europa Clipper gives us a clue to understand the time variability of the wave-particle acceleration and generation of high-energy electrons. In addition to that, if we could observe the change of synchrotron radiation spectrum around Europa orbit, it could be a signature of wave-particle acceleration. Therefore, time-spatial variation, as well as the spectral characteristics, are important to discuss the effect of wave-particle interaction.

Saturn also has a radiation belt. The magnetic field strength of Saturn is 1/10 of Jupiter, and the electron energy is less than 10 MeV due to the loss by the main ring. For these reasons, the intensity of Saturn's synchrotron radiation is much weaker than that of Jupiter. The expected total flux at 100-300MHz is 0.15-0.45mJy \citep{lorenzato2012}. The integration time ($\sim$ 1 hour) are needed for SKA 1-low to detect Saturn's synchrotron radiation. Observation of Saturn's synchrotron radiation was tried with LOFAR, but the synchrotron radiation from Saturn has not been found yet. If SKA 1-low can observe Saturn's radiation belt, which will give us more information on comparative studies on the planetary radiation belt.

\section{Thermal emissions: planetary atmosphere}
\label{sub:therpla}

\begin{figure}[b!]
\centering
\includegraphics[bb=0 0 524 515, width=0.46\hsize]{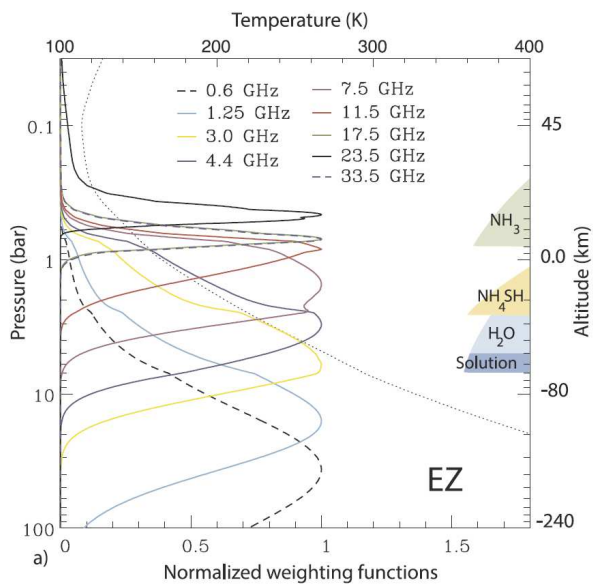}
\caption{Weighting functions of the Jovian thermal radiation at several wavelengths \citep{dePater2019}. A weighting function indicates to which altitudes the thermal radiation at the selected wavelength is sensitive. A thermochemical equilibrium atmospheric model for the Equatorial Zone is used for the calculation. A dotted line represents the atmospheric temperature (upper abscissa) and the locations of the cloud layers are also illustrated in the right ordinate.}\label{fig:jup_cf}
\end{figure}

\subsection{Radio emission from the deep atmospheres of the giant planets}
\label{ss:thermalatmos}
Maps of thermal emissions from the planetary atmosphere tell us the spatial distribution of atmospheric temperature and also the abundances of opacity sources (i.e., atmospheric compositions). 
In general, the observed thermal radiation originates from the atmosphere where the atmospheric optical depth (along the line-of-sight) becomes around unity. 
The atmospheric opacity varies with wavelength due to the absorption by molecules and clouds in the atmosphere. 
This means that different altitude levels of planetary atmospheres can be probed by selecting different observation wavelengths. 
Figure\,\ref{fig:jup_cf} is a good example to illustrate this point for the case of Jovian radio thermal emission. 
The plotted kernels, called weighting functions, are the derivatives of the atmospheric transmittance with respect to height, which indicate the weight of the thermal radiation from each altitude observed at a specific frequency. 
In Jupiter's atmosphere, a strong absorption due to NH$_3$ exists at 23--24 GHz. 
Selecting the observation frequencies lower than this NH$_3$ absorption frequency gives an access to a deeper level of the Jovian atmosphere.  
NH$_3$ is also the major absorbing gas in Saturn atmosphere. 
In Uranus and Neptune, H$_2$S increases its contribution to the atmospheric opacity at a wide frequency range of $\sim$1--10s GHz \citep{dePater1991}. 
As these species in the giant planets (NH$_3$ and PH$_3$ gases in Jupiter and Saturn and H$_2$S in Uranus and Neptune) have the source within the deep atmosphere, their 3D distributions can be used to constrain the ascending and descending transportation of the air from/to a deeper level.

Several scientific questions with respect to the giant planet atmosphere, which are expected to be addressed through future SKA observations, are reviewed by \citet{butler04}. 
Instead of repeating them in this paper, here we briefly describe some new findings obtained with VLA and the Juno explorer. 

VLA has been used several times for observing Jupiter (e.g., \citet{dePater2016}, \citet{dePater2019}) as well as other giant planets. 
Those VLA observations are often conducted with the data integration for several days, for typically $\sim$10 hours on each day, in order to accumulate the data with different array configurations. 
Therefore, the observed radio maps are smeared longitudinally. 
\citet{dePater2019} performed an intensive analysis on the VLA Jupiter maps at 3--37 GHz. 
From the longitude-smeared maps, the authors found the presence of a specific latitude range (8.5--11$^{\circ}$N) that contains a series of hot spots. 
\citet{dePater2019} also succeeded in reconstructing longitude-resolved maps via applying an innovative image synthesis technique, called as the ``facet technique'' which corrects the distortion of the image caused by the planetary rotation \citep{sault2004}. 
The longitude-resolved maps showed several hot spots and the ammonia plumes in Jupiter deep atmosphere. 
Some large plumes seem to be supersaturated with NH$_3$ gas at the pressure level up to 0.5 bar, indicating that the NH$_3$ gas is lifted above the main NH$_3$ cloud deck before being condensed out.
Further SKA observations with an improved spatial resolution and sensitivity will provide more detail visualization of those hot spots and plumes. 

There is also another scientific interest on the abundance of H$_2$O in Jupiter as it will help us to constrain the formation model of this gas giant.    
Recently, the Microwave Radiometer (MWR) onboard the Juno explorer has obtained H$_2$O abundance in the equatorial region at the pressure levels from 0.7 to 30 bar \citep{Li2020}. 
The derived H$_2$O abundance is $\sim$2.7 times enhanced compared to that anticipated from the protosolar O/H ratio. 
However, the spatial coverage of Juno/MWR observations is limited and it still remains unanswered whether this obtained H$_2$O abundance can be representative of the global abundance of Jovian atmosphere. 

\subsection{Radio emission from the terrestrial planet atmosphere}
\label{sub:terrestrialatmos}

Regarding to the terrestrial planets, low frequency radio observations have been mainly used for the surface science, while their atmospheres are more studied at higher frequencies (mm-wave or submm-wave). 
Having said that, Venus atmosphere can be an interesting target for a low frequency observation.

\subsubsection{Revealing the Venus atmosphere beneath the clouds}
\label{subsub:vensci}

Venus has a massive CO$_2$ atmosphere (90-bar pressure at the surface) and is globally covered with thick H$_2$SO$_4$ clouds laying at altitudes of 50--70 km. 
It is well-known that the Venusian atmosphere rapidly rotates the planet, a.k.a. ``super-rotation'' of the atmosphere, reaching to the speed $\sim$60 times faster ($>$100 m/s at the cloud-top altitude) than the planetary rotation. 
To maintain such a global and rapid atmospheric circulation, the angular momentum of the planetary rotation should be transported upwards to the atmosphere and distributed over all the latitudes. 
This maintenance mechanism has been a long-standing unsolved question in Venus atmospheric science.  
Interestingly, super-rotating atmospheres are also found in Titan, Jupiter, and Saturn. 
Therefore, the understanding of super-rotation of atmospheres in the solar system can be an essential subject in a broader field of the planetary science \citep{imamura20}.

Recently, a part of the mechanism of the super-rotation has been finally revealed by \citet{Horinouchi2020} who analyzed the wind velocity maps obtained by the Japanese Venus orbiter Akatsuki. 
The authors concluded that the super-rotation at the equatorial region is maintained by the atmospheric thermal tides which are excited by the contrast of solar heating between day and night. 
They also pointed out that the angular momentum at the equator is transported to middle latitudes by meridional circulations. 
Despite this remarkable discovery, the full mechanism of the super-rotation still remains unrevealed.   
This is because the altitude levels probed by Akatsuki are limited around the cloud layer, and therefore we are still unable to constrain the angular momentum transport between the surface and the sub-cloud level. 

\begin{figure}[bt]
\centering
\includegraphics[bb=0 0 489.7 374.5, width=0.53\hsize ]{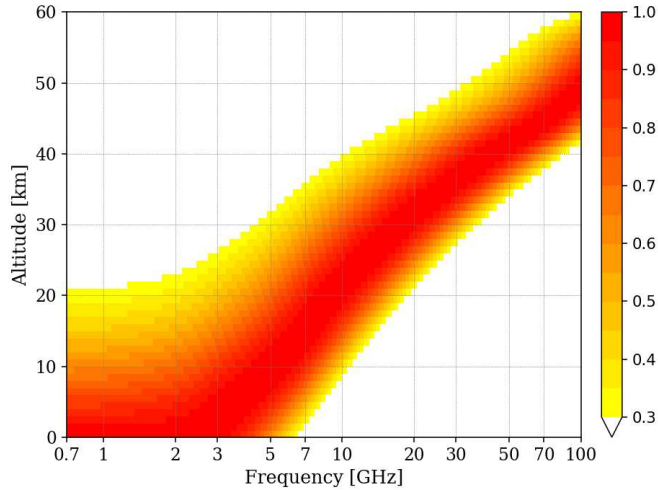}
\caption{
Weighting function of the radio radiation from Venus atmosphere. 
A standard atmospheric condition and the gas opacities due to CO$_2$ and SO$_2$ are assumed, whereas the thermal emission from the sub-surface layers are not considered (which should have some contribution at a frequency below $\sim$1 GHz).   
}\label{fig:venus}
\end{figure}

The Venusian clouds are so thick that almost no optical/infrared wavelengths can penetrate the atmosphere beneath the clouds. 
There are only two atmospheric windows enabling us to remotely observe the Venus atmosphere beneath the clouds: at near-infrared region and at radio frequencies. 
The near-infrared atmospheric windows are used to study the spatial distribution of trace gas (such as CO, H$_2$O, SO$_2$) below the clouds. However it is not possible to derive the temperature information from these near-infrared observations.  
On the other hand, the Venus atmospheric opacity at radio frequencies is dominated by the Collision Induced Absorption (CIA) of CO$_2$ \citep{Gruszka1997}. 
The absorption coefficient of CO$_2$ CIA becomes smaller at larger wavelengths, i.e., the atmosphere becomes more transparent at larger wavelengths.
The second significant absorbing gas is SO$_2$. 
The absorption lines of SO$_2$ at radio frequencies are not so strong, but there exist plenty of lines. The width of each line becomes very broad due to the high atmospheric pressure and affects like a continuum absorption. 
Figure \ref{fig:venus} shows the weighting functions for low frequency radio observations of Venus atmosphere. 
It clearly illustrates that the thermal radiation emitted from the atmosphere at altitudes below $\sim$30 km, which is significantly below the cloud layer, is detectable at radio frequencies. 

\begin{figure}[bt]
\centering
\includegraphics[bb=0 0 1070 361, width=0.88\hsize ]{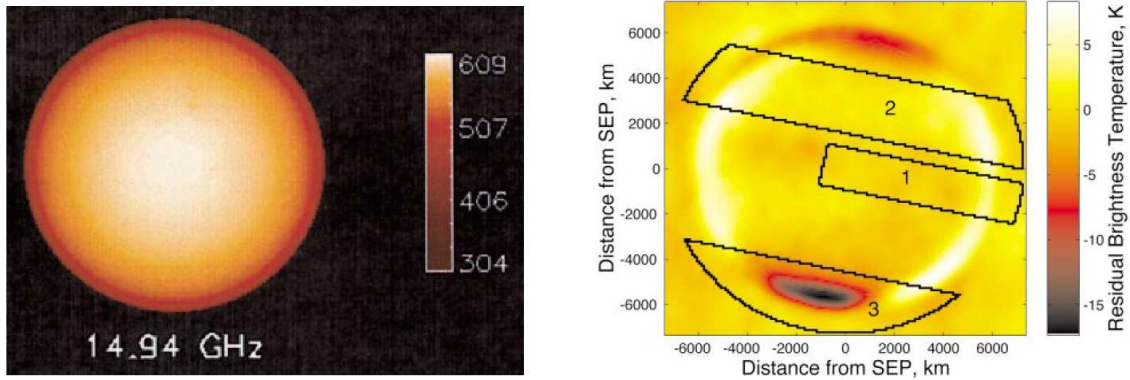}
\caption{
Left panel: Venus radio map at 14.94 GHz obtained with VLA \citep{Butler2001}. Right: Residual image obtained after subtracting the disk-averaged brightness temperature \citep{Jenkins2002}.
The regions 1--3 represent the area where \citet{Jenkins2002} analyzed in their paper. 
}\label{fig:ven_vla}
\end{figure}

\citet{Jenkins2002} conducted Venus observations using VLA and obtained the thermal map probing the altitude of 20--30 km (Fig.\,\ref{fig:ven_vla}). 
The authors pointed out that the temperature at that altitude is warmer by $\sim$25 K than that of the Venus standard atmosphere model (a model compiled from the past \textit{in-situ} measurements of the Pioneer Venus mission). 
It is noted that this VLA analysis was done in a global average and also the SO$_2$ abundance was assumed to be 150 ppm when deriving the temperature distribution. 
Improving the observation sensitivity so that we can map the spatial distribution of the thermal emission and preparing more accurate knowledge on the SO$_2$ abundance will bring a breakthrough to understand the 3D structure of Venus atmospheric temperature beneath the cloud layer. 

Temperature distribution at the lowermost atmosphere of Venus remains largely unobserved. 
Only a few \textit{in-situ} measurements by past entry probes and landers were available.   
The measurement by the VeGa-2 descending probe suggested that a convective layer appears at altitudes from 18 to 32 km, but the nature of such a thermal structure is still not understood clearly \citep{Lebonnois2017}. 
Another puzzling outcome of the VeGa-2 probe was the presence of highly unstable layer (the atmospheric static stability was derived to be as negative as $-$1.5 K/km) at altitudes below 7 km. 
To interpret this VeGa-2 measurement as a ``stable'' atmospheric condition, one needs to dare to presume that the mean molecular mass of the atmosphere (atmospheric composition) has a vertical gradient (which is never observed in the Earth's well-mixed troposphere).  
\citet{Lebonnois2017} proposed a possibility of density-driven separation of N$_2$ from CO$_2$ at that unstable layer perhaps due to molecular diffusion or natural density-driven convection. 
New sensitive measurements of the vertical profile of the temperature near the surface will test this hypothesis. 

Detecting temperature disturbances due to atmospheric waves and tides can be a more challenging target because of the small amplitude ($\sim$0.05 K against the background atmospheric temperature of $\sim$500--700 K), but it is the most desired information because such wave activities are responsible for distributing the angular momentum over the latitudes and altitudes.

\subsubsection{Detectability: SKA1-MID}
\label{subsub:venska}

The key questions that SKA will address with respect to the Venus atmosphere are as follows. 
\textbf{
\begin{itemize}
    \item What is the vertical structure of the atmospheric temperature in the lower atmosphere? In particular, how stable is the near-surface atmosphere? 
    \item What is the spatial and temporal variability of the temperature structure beneath the optically thick clouds?
    What kinds of atmospheric waves and tides are generated there?
\end{itemize}
}




As already shown in Fig.~\ref{fig:venus}, the thermal radiation from the lowermost atmosphere can be detected by using SKA, particularly at the highest frequencies of SKA1-MID.
In this frequency, SO$_2$ in the lower atmosphere also contributes to the atmospheric opacity as well as CO$_2$ CIA. 
The spatial distribution of atmospheric temperature can be retrieved from a radio brightness map by making use of complementary near-infrared observations which constrain the abundance of SO$_2$. 
We summarize the required observation capability for SKA as follows. 
\begin{itemize}
\item \textit{Measurements of vertical profile of the temperature.} 
In order to understand the meteorological behaviour, it is important to obtain the lowermost atmospheric temperature profiles at different latitudinal regions (equatorial, mid-latitudes, and polar region) and at different local times (day and night). Venus apparent angular diameter varies from 10 to 60 arcsec. The spatial resolution of SKA1-MID will be better than $\sim$1 arcsec at the frequency higher than $\sim$1 GHz \citep{skatechdoc}. This is sufficient to resolve the different latitudes ($\sim$10s$^{\circ}$) of Venus even when it has the smallest apparent angular diameter. 

The vertical resolution of SKA observation is expected to be $\sim$10-15 km (roughly estimated from the half-width of the altitude profile of the contribution function in Fig.~\ref{fig:venus}). 
This capability is sufficient to vertically resolve the distinct layers found by the VeGa-2 probe (namely, the highly unstable, stable, and unstable layers from 0--7 km, 7--18 km, and 18--32 km, respectively). 

The SKA1-MID achieves a $\mu$Jy/beam-level sensitivity for the continuum emission (with 1 hr observation) \citep{skatechdoc}. This corresponds to the sensitivity of a few Kelvin in brightness temperature of $\sim$1 arcsec$^2$ beam at 1 GHz, which is sufficient to constrain the vertical lapse rate of the atmospheric temperature (the adiabatic lapse rate is 7.6 K/km in Venus).

\item \textit{Detection of thermal disturbances due to atmospheric waves and tides.} 
It is difficult to estimate the detectability of thermal disturbances because there is no reference information available in past observations. 
Some numerical studies predicted the presence of thermal tides beneath the cloud layer with an amplitude of $\sim$0.05 K (\citet{Takagi2018}, \citet{Lebonnois2016}). 
This means that a mK-level sensitivity is required for the detection. 
The thermal tide should have an east-west wavenumber of one as the most dominant component, i.e., the longitudinal (diurnal) distribution of the temperature has one local maximum and one local minimum along the planet. 
Therefore we do not need a very high spatial resolution for detecting such a longitudinal temperature variation: a beam size of $\sim$5-10 arcsec can be accepted. 
With this relaxation of the spatial resolution, we anticipate that the proposed observation is feasible with SKA1-MID.  
\end{itemize}


\section{Thermal emissions: icy bodies}
\label{ss:icy_body}

\subsection{Recent observations of icy bodies}
\label{ss:icy_body_obs}

For an airless object (including icy moons with tenuous atmosphere), the thermal radiation comes from the sub-surface layer with a penetration depth of $\sim$10--20-fold the wavelength. 
The penetration depth depends on the physical property of the surface material. 
The brightness temperature $T_B (\lambda)$ at a wavelength $\lambda$ can be expressed with the radiative transfer equation 
\begin{equation}\label{eq:subsurf1} \displaystyle
   T_B (\lambda) = \left(1 - R(\lambda, \phi) \right) \int_{0}^{\infty} T(z) \exp \left( \frac{- \kappa_{\lambda} z}{\cos \theta} \right) \frac{\kappa_{\lambda}}{\cos \theta} dz,
\end{equation}
where $R (\lambda , \phi)$ is the Fresnel reflectivity coefficient for an emission angle $\phi$ thus $1 - R$ represents the emissivity. 
$T (z)$ is the physical temperature at a depth $z$ below the surface, $\kappa_\lambda$ is the radio absorption coefficient, and $\theta$ is an angle at which the radiation propagates below the surface. 
The relationship between $\phi$ and $\theta$ is described by Snell's refraction law as 
\begin{equation}\label{eq:subsurf2} \displaystyle
\sqrt{\epsilon} = \frac{\sin\theta}{\sin\phi}
\end{equation}
where $\epsilon$ is the complex dielectric constant. 
The radiation is polarized when it is emitted from the surface. 
The degree of polarization is a function of $\epsilon$, $\theta$, and $\phi$. 
Therefore, by observing the parallel and perpendicular polarized radiation under different emission angles, we can retrieve information about $T(z)$ and $\epsilon$. 
The dielectric constant can be used to constrain the physical parameters of the surface (materials, surface roughness, porosity, etc.). 

Thermal emission from the icy body's surface is likely associated with the geological activities in the icy crust and water plume from the interior ocean. 
In particular, the icy moons at the gas giants plumes the interior ocean water accompanying the tidal heating by the orbital motion. 
The interior ocean of Saturn's moon Enceladus has already been confirmed by the \textit{in-situ} measurements of the water plume with the Cassini explorer. 
Some precursor materials of the amino acid have also been identified in the Enceladus water plume with the dust counter and ion mass spectrometer onboard Cassini \citep{khawaja19}. 
This suggests that the interior oceans of icy moons at the gas giants are highly possible habitable environments.

\begin{figure}[tb!]
\centering
\includegraphics[width=1.\hsize]{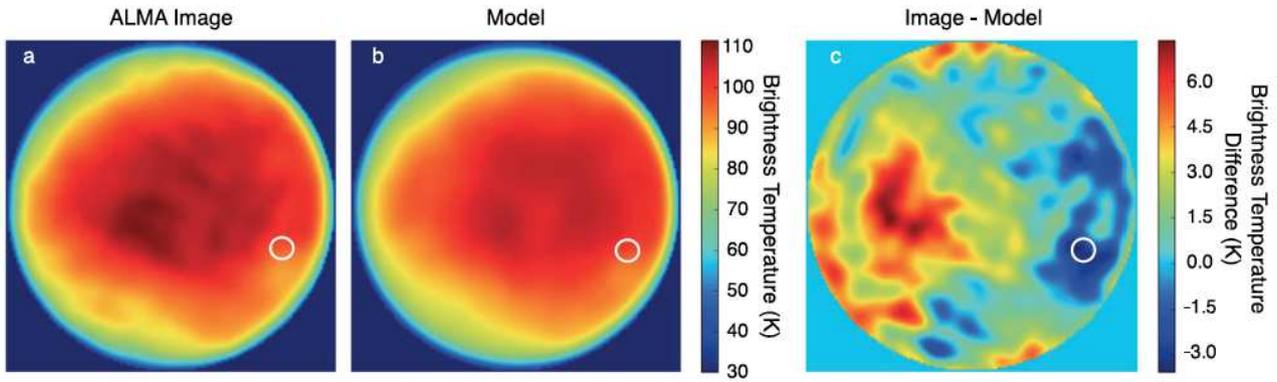}
\caption{Distribution of Europa's surface brightness temperature measured with ALMA \citep{trumbo17}. 
Left: Brightness temperature taken from the ALMA observation. Center: that from the thermal conduction model. Right: the difference between the observation and the model. 
A spatially uniform thermal inertia is given in the thermal conduction modeling. 
The significant difference between the observation and model suggests the non-uniformity in the thermal inertia. 
A circle with the white solid line indicates the beam size of ALMA at the observing time, which is located at the possible water plume location suggested from the previous Galileo infrared imaging. \label{fig:trumbo_18}
}
\end{figure}

The water plume and crack at the bottom of the plume are expected to have higher temperature compared to other geologically inactive regions \citep{spencer99}. 
\citet{trumbo17} and \citet{trumbo18} measured the surface distribution of Europa's thermal radiation by using the ALMA radio telescope (Figure \ref{fig:trumbo_18}). 
They did not confirmed the significant enhancement of thermal emission in the region that likely has the water plume expected from the previous infrared observations with the Galileo explorer \citep{spencer99}. 
They interpreted that the absence of thermal emission is due to its temporal variation. 
It was concluded that the thermal emission distribution can be explained by  modeling of the surface heat conduction taking account of spatial inhomogeneity of the surface albedo, solar radiation flux, and surface thermal inertia. 
Specifically, the thermal inertia distribution is likely associated with the oceanic water that erupted and accumulated on the surface \citep{trumbo17,trumbo18}. 
Although the cause of the non-uniform thermal inertia is unknown, it is likely due to differences in the crystal structure, porosity, and particle size of the fresh surface material newly created by the sedimentation of erupted water.

The \textit{in-situ} plasma wave measurements with the Galileo explorer during the Europa flyby suggested that the water plume has a spatial extent of 100s km at an altitude of a few 100 km \citep{jia18}.
This implies the spatial scales of water plume and footprint crack of a few 100 km or less, which may not be resolved with the ALMA spatial resolution ($\sim 200$ km at Europa). 
The thermal radiation imaging obtained with a spatial resolution better than ALMA will resolve the thermal inertia distribution including the water plume structure, which contributes to understanding of the surface material environments.

The timescale of the water plume is a big issue to be resolved. 
\citet{roth14} first discovered the appearance and disappearance of a possible water plume based on the UV hydrogen and oxygen imaging with the Hubble Space Telescope. 
It would be reasonable to expect that the water plume occurs at a specific phase out of 85-hr orbital period because of the tidal force that depends on orbital motion.
\citet{roth16} made a long-term monitoring for the Europa water plume at each orbital phase to demonstrate the orbital phase dependence. 
However, they and following observations did not find such dependence.
This suggests that the change in the tidal force associated with the orbital motion is not enough for the water plume eruption, or the number of observations at each orbital phase is not yet sufficient to detect it. 
Another possible cause for the absence of orbital phase dependence is the contamination by Jupiter's magnetospheric electrons that excites the hydrogen and oxygen UV emission via collision with Europa's water plume gas. 
This collisional UV excitation likely mimics the temporal variability in the water plume eruption. 
The highly-resolved radio interferometry imaging for the water plume thermal emission will reasonably identify the orbital phase dependence of water plume as a complementary observation method.

\subsection{Imaging of thermal emissions from the icy bodies with SKA}
\label{ss:thermal_SKA}

We can extract some simple key questions for the icy bodies from the reviews in \S\ref{ss:icy_body_obs}:
\textbf{
\begin{itemize}
    \item What are the spatial scales of the water plume and the footprint crack?
    \item What are the temporal scales of the water plume and the associated geological activities of the crust and ocean?
\end{itemize}
}

The spatial distribution and occurrence rate of the water plume are still unknown because of insufficient sensitivity and resolution of the current observations.
Therefore, the geological activities in the interior ocean and icy crust are not well understood although these are parts of the extra-terrestrial habitable environments. 
The high resolution and high sensitivity observation of the thermal emission with SKA visualizes the water plume, footpoint crack, and their dynamics. 
By the SKA observation, we may evaluate the loss rate of the interior water ocean and amino acid precursors, which contributes to understanding persistence of the interior ocean's habitability.
Similar SKA observations are applicable to other tenuous atmospheric bodies to visualize their geological activities that likely constrains the surface and interior evolution. 
Here we briefly make detectability studies of thermal emissions from the icy bodies for SKA1 and 2.

\subsubsection{Detectability: SKA1} 

\begin{figure}[htb]
\centering
\includegraphics[width=0.7\hsize]{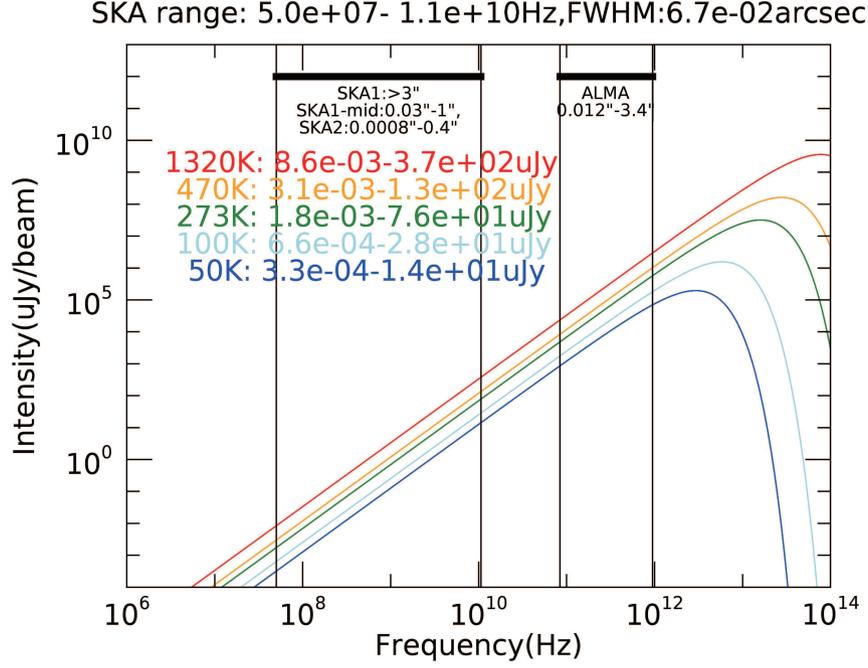}
\caption{
Black body spectrum with surface temperatures at our solar system bodies. 
Here we assumed that the black body emission source uniformly distributes within the beam size FWHM of 67~mas. 
Each colored solid line corresponds to the surface temperature: (red) Io's volcano with a high temperature of 1320~K, (blue) Io's volcano with a low temperature of 470~K, (green) water or ice at 273~K, (light blue) sunlight region of Jupiter's icy moon, (blue) Kuiper belt object at 50K. The range of radiation intensity at the SKA frequency range is labelled with the colored text. }\label{fig:thermal}
\end{figure}

Figure \ref{fig:thermal} presents the estimate for the surface thermal emission approximated by black body spectra with varying temperatures.
Here we assume the SKA1 observation for a uniformly distributed radio source within a beam FWHM of 67 milliarcseconds (mas) \citep{skatechdoc} at 10.61 GHz, where the icy moon's thermal emission is likely maximum in the frequency range observable with SKA.
The black body radio flux ranges from $3.3\times 10^{-4} \mu$Jy beam$^{-1}$ (50MHz) to $14 \mu$Jy beam$^{-1}$ (10.61GHz). 
The detection limit of SKA1 with the 1 hr exposure time is $1.2 \mu$Jy beam$^{-1}$  \citep{skatechdoc}, which corresponds to a significant S/N ratio of 12$\sigma$ at 10.61GHz. 
The apparent diameter of Europa around the opposition is 1~arcsec, which is separated into 15 pixels with the SKA1 beam size of 67~mas, equivalent to a few 100 km on the surface. 
This resolution is comparable to the ALMA resolution of 50~mas at 230 GHz. 
As shown in Figure \ref{fig:thermal}, the thermal emission intensity at the SKA frequency range is a few orders of magnitude or more less than those at the ALMA range.
However, the high sensitivity observation with SKA1 compensates the large difference in the source intensity between the SKA1 and ALMA.
The S/N ratio of SKA1 is just one order of magnitude less than ALMA.

\subsubsection{Detectability: SKA2}

The SKA2 will achieve the sensitivity comparable to ALMA and the high spatial resolution 10 times greater than ALMA ($\sim$1~mas a few km on the surface). 
With the high sensitivity and resolution, SKA2 would detect the water plume resolving the plume and crack sizes ($<$ a few 100km) that was not able to be made with ALMA. 
The difference of thermal emission intensity between the water plume and geologically inactive region is unknown.
However, the amplitude of thermal emission associated with the inhomogeneity in the thermal inertia was suggested to be 10\% \citep{trumbo17, trumbo18}.
SKA2 can detect the 10\% amplitude with S/N$>10$ by an one-hour or longer exposures. 
The thermal emission inhomogeneity at each phase of Europa's 85-hour orbital period would unveil the geological activities associated with the interior ocean driven by the tidal force of orbital motion.
This observation has not been made even with HST or ALMA. 

\subsubsection{Synergy with future explorations}

In the early 2030s, the JUICE explorer (ESA) and Europa Clipper (NASA) are going to make the \textit{in-situ} measurements of the environments of Jupiter's icy moons Europa and Ganymede by the flyby and orbital exploration.
The surface and water plume of the icy moons will be explored in detail through the surface imaging, plasma and field measurements, and so on with JUICE and Europa Clipper. 
However, the spacecraft makes their observations only at its location, with which it is hard in principle to resolve the spatio-temporal structures of the icy moon's environment that dynamically varies and interacts with Jupiter's magnetosphere.
SKA, which is planned to operate from late 2020, is one of the most effective facility that will globally visualize the environment, which complementarily gives the context of environmental parameters around the spacecraft location.
For example, SKA thermal emission imaging will detect a temporal appearance of localized hot region on the surface, of which molecular emission, electron density, and ion composition are quantitatively explored by the \textit{in-situ} measurements in the high resolution mode with the spacecraft.
Ultraviolet emission from the neutral oxygen and hydrogen tenuous atmosphere associated with the water plume activity is going to be continuously monitored from Earth's orbit with the LAPYUTA space telescope, which is now being proposed as a future mission led by some of the authors of this paper with Japanese planetary science community.
The synergy between SKA, JUICE, Europa Clipper, and LAPYUTA will surely unveil the current dynamics of icy moon's environment including the interior ocean.

\section{Summary and future perspectives}
\label{s:summary}

\subsection{Non-thermal emissions: auroral radio emissions}
We reviewed the physical characteristics of non-thermal auroral radio emission at the solar system planets in \S\ref{s:auroral}, with emphasis on the gas giants and their moons. 
The auroral radio emission is excited via the plasma instability in the auroral electron velocity distribution CMI at an emission frequency of $\omega \sim \omega_c$ with anisotropic beaming of $10\mathrm{s}^\circ$ with respect to the background magnetic field line (\S\ref{ss:auroral_generation}).
The variety of energy source processes for the auroral radio emission have been proposed: the Hill current system (\S\ref{sss:auroral_current_hill}), Io-Jupiter current system (\S\ref{sss:auroral_current_io}), and solar wind (\S\ref{sss:auroral_current_sw}). 
Note that the Hill and Io-Jupiter current systems are driven by the planetary rotation, intrinsic magnetic field, atmosphere (ionosphere), and corotating magnetospheric plasma originating from the moons. 
The radio energy sources are modified by the transfer process of the ``moongenic'' plasma in the magnetosphere (\S\ref{sss:mass_transfer}) and by the solar EUV flux irradiated to the polar atmosphere (\S\ref{sss:solar_euv}). The Alfv\'{e}nic acceleration was recently found to be dominant for Jupiter's auroral acceleration from the keV to MeV energy range by the Juno's \textit{in-situ} measurements above the auroral region. We briefly studied detectability of the newly found Alfv\'{e}nic acceleration process for the future SKA1\&2 observations, which was found to be significant to be detected with SKA1 and imaged by second-by-second monitoring (\S\ref{ss:detectability}).

The physics of the auroral radio emission reviewed here for the solar system planets are potentially applicable to exploration for exoplanets. 
Based on these physics, essential environments of exoplanet will be explored by the future observations with SKA. 
Detection of the auroral radio emission from an exoplanet strongly demonstrates its atmosphere because the auroral current system requires the finite electric ionospheric conductivity at the planetary atmosphere (\S\ref{sss:auroral_current_hill}). 
Spectrum of the exoplanetary radio emission constrains the magnetic flux density at the planetary surface with the CMI emission frequency $\omega \sim \omega_c$ (\S\ref{ss:auroral_generation}).
Rotation period of the exoplanet is estimated from the periodicity of auroral radio emission as made for Jupiter's rotation period (\cite{dessler02}).
Periodic occurrence of auroral radio emission in a particular phase relationship with the planetary rotation period indicates the Io-Jupiter like auroral current system, 
which is demonstration of the atmosphere of ``exomoon'' because it is required as the finite electric conductivity region to maintain the current system (\S\ref{sss:auroral_current_io}).
Transient auroral radio bursts with a period of 2--3 days are suggestive of the plasmoid mass ejection from the exoplanetary magnetosphere.
That also indicates that the exomoons erupt the mass comprising the volcanic and/or water gases to the magnetosphere (\S\ref{sss:mass_transfer}).
Simultaneous observation of a host star with its exoplanets based on the optical and radio observations could visualize propagation of the coronal-mass ejection (CME) and other stellar wind shock structures in the stellar system: i.e., the exoplanets work as ``sensors'' for the stellar wind that show intensification in the auroral radio emission following the stellar wind shock arrival (\S\ref{sss:auroral_current_sw}, \cite{gurnett02, prange04}). 
Energy inputs from the host star to the exoplanetary magnetospheres are estimated from the auroral radio intensity in combination with the scaling-law  proposed by \citet{zarka07} (\S\ref{sss:auroral_current_sw}, Figure \ref{fig:zarka_07}).

\subsection{Non-thermal emissions: synchrotron radiation}
Here we reviewed the physical processes in Jupiter's radiation belt (\S\ref{ss:radiationbelt}). High-sensitive and wide-area imaging of Jupiter's synchrotron radiation will be possible using SKA. The area extends around Europa orbit at 50 MHz and Io at 325MHz. SKA can reveal the physical characteristics of the three acceleration processes in Jupiter's radiation belt: pre-heating, betatron acceleration, and the wave-particle interaction. Combining \textit{in-situ} observation of JUICE and Europa Clipper with SKA observations is also important to achieve our goals (\S\ref{ss:obs_jsr}). 

It is also possible to extend this study to exoplanet radiation belts. Magnetized planets have radiation belts, and the relativistic electrons in the radiation belts emit synchrotron radiation. Analogous to the radiation belts in the solar system, they should vary with the activity of the host star. 
The diversity of the synchrotron radiation among differnet planetary systems provides an important evidence for understanding the fundamental physical process in the radiation belt. The problem is the intensity of exoplanet synchrotron radiation is by far weaker than that of the auroral emission. The full performance of SKA is not likely to be sensitive enough to observe the exoplanets' synchrotron radiation.

\subsection{Thermal emissions: atmospheres}

Radio thermal emission observation is a powerful tool to investigate the deep levels of the giant planet atmosphere \citep{butler04}. 
In addition, Venus is also an important science target to probe its atmosphere. 
The frequency covered by SKA1-MID will provide a unique access to the lowermost altitude of Venus atmosphere (\S\ref{subsub:vensci}). 
The vertical temperature profile from the surface to $\sim$30 km altitude can be constrained by SKA1-MID, which will be used to understand how stable/unstable the deep atmosphere is. 
The hypothesis of density-driven separation of N$_2$ at the near surface \citep{Lebonnois2017} will be investigated.  
Moreover, the high sensitivity of SKA will detect the thermal tides propagating at the atmosphere beneath the clouds. 
The activities of such tides (including other atmospheric waves) are crucial for understanding the angular momentum transportation within the Venus atmosphere, i.e., acceleration and deceleration of atmospheric circulation (\S\ref{subsub:venska}).  
Understanding the structure, stability, and circulation of the Venusian atmosphere will also develop our understanding to tidally-locked exoplanets.

\subsection{Thermal emissions: icy bodies}

The interior ocean of the icy moons of the gas giants is the best candidates of extra-terrestrial habitable environment in our solar system.
The icy moon, particularly Europa orbiting Jupiter, has been observed with the space UV telescope HST and ground-based radio telescope ALMA (\S\ref{ss:icy_body_obs}).
Although the HST UV imaging detected the expansion of the tenuous oxygen atmosphere possibly associated with the plume of water originating from the interior ocean \citep{roth14, roth16}, 
there had not been the detection of thermal emissions from the water plume and the footprint crack with ALMA and other observation facilities \citep{trumbo17, trumbo18}. 
The future imaging with SKA2, achieving as high spatial resolution as $\sim$1~mas (a few km on the surface), will resolve the localized thermal emissions of water plumes and cracks with a size of $<$ a few 100 km \citep{jia18} (\S\ref{ss:thermal_SKA}).
The thermal inertia inhomogeneity of the icy moon's surface was suggested to be 10\% \citep{trumbo17, trumbo18}, which is detectable for SKA2 with S/N$>10$ by an one-hour or longer exposure. 
The coordinated observations of SKA with the future explorers and earth-based telescopes like JUICE, Europa Clipper, and LAPYUTA will uncover the distribution and dynamics of water plumes and interior oceans.


\newpage

\begin{thebibliography}{}






\bibitem[Bagenal \& Delamere (2011)]{bagenal11}
Bagenal, F., \& Delamere, P.~A. 2011, Journal of Geophysical Research: Space
 Physics, 116






\bibitem[Bolton et~al.(2002)]{bolton2002}
{Bolton}, S.~J., et~al. 2002, Nature, 415, 987




 

\bibitem[Braun et~al.(2017)]{skatechdoc}
Braun, R., and SKA Science Team, 2017, SKA-TEL-SKO-0000818

\bibitem[Brice \& McDonough (1973)]{brice1973}
Brice \& McDonough (1973) Jupiter's radiation belts Icarus, 18, doi:10.1016/0019-1035(73)90204-2




\bibitem[Butler et~al.(2001)]{Butler2001}
{Butler}, B.~J., {Steffes}, P.~G., {Suleiman}, S.~H., {Kolodner}, M.~A., \&
 {Jenkins}, J.~M. 2001, Icarus, 154, 226

\bibitem[Butler et~al.(2004)]{butler04}
Butler,B., D.~Campbell, I.~de~Pater, and D.~Gary, 2004, New Astronomy Reviews, 48, 11-12, 1511.







\bibitem[Clarke et~al.(2004)]{clarke04}
Clarke, J.~T., Grodent, D., Cowley, S.~W., Bunce, E.~J., Zarka, P., Connerney,
 J.~E., \& Satoh, T. 2004, Jupiter: The Planet, Satellites and Magnetosphere,
 1, 639

\bibitem[Clark et~al.(2017)]{clark17}
Clark, G., et al. (2017), Energetic particle signatures of magnetic field-aligned potentials over Jupiter's polar regions, Geophys. Res. Lett., 44, 8703– 8711, doi:10.1002/2017GL074366.


\bibitem[Connerney et~al.(2018)]{connerney18}
Connerney, J., et~al. 2018, Geophysical Research Letters, 45, 2590


\bibitem[Cowley and Bunce (2001)]{cowley01}
Cowley, S. and E.~Bunce, 2001, \emph{Planetary and Space Science},  49, 10-11, 1067.




\bibitem[Damiano et~al.(2019)]{damiano19}
Damiano,P., P.~Delamere, B.~Stauffer, C.-S. Ng, and J.~Johnson, 2019, \emph{Geophysical Research Letters}, 46, 6, 3043.


\bibitem[de Pater \& Goertz (2018)]{depater1990}
de Pater, I., and Goertz, C. K. (1990), Radial diffusion models of energetic electrons and Jupiter's synchrotron radiation: 1. Steady state solution, J. Geophys. Res., 95( A1), 39– 50, doi:10.1029/JA095iA01p00039.


\bibitem[de Pater et~al.(1991)]{dePater1991} 
{de Pater}, I., {Romani}, P.~N., \& {Atreya}, S.~K. 1991, Icarus, 91, 220

\bibitem[de Pater et~al.(2016)]{dePater2016}
{de Pater}, I., {Sault}, R.~J., {Butler}, B., {DeBoer}, D., \& {Wong}, M.~H. 2016, Science, 352, 1198

\bibitem[de Pater et~al.(2019)]{dePater2019} 
{de Pater}, I., {Sault}, R.~J., {Michael}, H.~W., {Leigh}, N.~F., {DeBoer}, D. \& {Butler}, B. 2019, Icarus, 322, 168


\bibitem[Dessler (2002)]{dessler02}
Dessler, A.~J. 2002, Physics of the Jovian magnetosphere, Vol.~3 (Cambridge
 University Press)


\bibitem[Ebert et~al.(2017)]{ebert17}
Ebert, R. W., et al. (2017). Spatial distribution and properties of 0.1–100 keV electrons in Jupiter's polar auroral region. Geophys. Res. Lett., 44. doi:10.1002/2017GL075106




\bibitem[Garrett et~al.(2005)]{garrett2005}
{Garrett}, H.~B., {Levin}, S.~M., {Bolton}, S.~J., {Evans}, R.~W., \&
 {Bhattacharya}, B. 2005, \grl, 32, L04104


\bibitem[Goertz et~al.(1979)]{goertz1979}
Goertz, C. K., Van Allen, J. A., and Thomsen, M. F. (1979), Further observational support for the lossy radial diffusion model of the inner Jovian magnetosphere, J. Geophys. Res., 84( A1), 87– 92, doi:10.1029/JA084iA01p00087.







\bibitem[Grima et~al.(2012)]{Grima2012}
{Grima}, C., et~al. 2012, Icarus, 220, 84

\bibitem[Gruszka \& Borysow (1997)]{Gruszka1997}
{Gruszka}, M., \& {Borysow}, A. 1997, Icarus, 129, 172

\bibitem[Gurnett et~al.(2002)]{gurnett02}
Gurnett, D., et~al. 2002, Nature, 415, 985





\bibitem[Han et~al.(2018)]{han2018}
{Han}, S., {Murakami}, G., {Kita}, H., {Tsuchiya}, F., {Tao},
 C., {Misawa}, H. et~al. 2018, Journal of Geophysical Research: Space Physics, 123, 9508-9516.

\bibitem[Hess et~al.(2008)]{hess08}
Hess, S., Cecconi, B., \& Zarka, P. 2008, Geophysical Research Letters, 35

\bibitem[Hill (1979)]{hill79}
Hill, T. 1979, Journal of Geophysical Research: Space Physics, 84, 6554

\bibitem[Hill (2001)]{hill01}
Hill, T. 2001, Journal of Geophysical Research: Space Physics, 106, 8101

\bibitem[Horinouchi et~al.(2020)]{Horinouchi2020} 
{Horinouchi}, T. et~al. 2020, Science, 368, 405.

\bibitem[Horne et~al.(2008)]{Horne2008}
{Horne}, R.~B., {Thorne}, R.~M., {Glauert}, S., {Menietti}, D., {Shprits},
 Y.~Y., \& {Gurnett}, D.~A. 2008, Nature physics, 4, 301






\bibitem[Imamura et~al.(2020)]{imamura20}
{Imamura}, T., {Mitchell}, J., {Lebonnois}, S., {Kaspi}, Y., {Showman}, A.~P., \& {Korablev}, O. 2020, Space Sci Rev, 216, 87


\bibitem[Itoh, (2008)]{itoh08}
Itoh 2008, Master Thesis of Tohoku University










\bibitem[Jenkins et~al.(2002)]{Jenkins2002}
{Jenkins}, J.~M., {Kolodner}, M.~A., {Butler}, B.~J., {Suleiman}, S.~H., \&
 {Steffes}, P.~G. 2002, Icarus, 158, 312


\bibitem[Jia et~al.(2012)]{jia12}
Jia, X., M. G. Kivelson, and T. I. Gombosi, 2012, Journal of Geophysical Research: Space Physics, 117, A04215.

\bibitem[Jia et~al.(2018)]{jia18}
Jia, X., Kivelson, M.~G., Khurana, K.~K., \& Kurth, W.~S. 2018, Nature
 Astronomy, 2, 459







\bibitem[Katoh et~al.(2011)]{katoh2011}
{Katoh}, Y., et~al. 2011, Journal of Geophysical Research (Space Physics), 116,
 A02215

\bibitem[Khawaja et~al.(2019)]{khawaja19}
Khawaja, N., et~al. 2019, Monthly Notices of the Royal Astronomical Society, 489, 5231


\bibitem[Kimura et~al.(2018)]{kimura18}
Kimura, T., et~al. 2018, Journal of Geophysical Research: Space Physics, 123,
 1885

\bibitem[Kimura et~al.(2013)]{kimura13}
Kimura, T., et~al. 2013, Journal of Geophysical Research: Space Physics, 118,
 7019

\bibitem[Kimura et~al.(2011)]{kimura11b}
Kimura, T., Tsuchiya, F., Misawa, H., Morioka, A., \& Nishimura, Y. 2011,
 Journal of Geophysical Research: Space Physics, 116

\bibitem[Kita et~al.(2015)]{kita2015}
{Kita}, H., et~al. 2015, Journal of Geophysical Research (Space Physics), 120,
 6614

\bibitem[Kita et~al.(2013)]{kita2013}
{Kita}, H., {Misawa}, H., {Tsuchiya}, F., {Tao}, C., \& {Morioka}, A. 2013,
 Journal of Geophysical Research (Space Physics), 118, 6106




\bibitem[Kollmann et~al.(2018)]{kollmann2018}
Kollmann, P., Roussos, E., Paranicas, C., Woodfield, E. E., Mauk, B. H., Clark, G., et al. (2018). Electron acceleration to MeV energies at Jupiter and Saturn. Journal of Geophysical Research: Space Physics, 123, 9110– 9129. https://doi.org/10.1029/2018JA025665



\bibitem[Lamy et~al.(2008)]{lamy08}
Lamy, L., \emph{et~al.}, 2008, Journal of Geophysical Research: Space Physics, 113, A07201.








\bibitem[Lebonnois et~al.(2016)]{Lebonnois2016} 
{Lebonnois}, S., {Sugimoto}, N., \& {Gilli}, G. 2016, Icarus, 278, 38.

\bibitem[Lebonnois \& Schubert (2017)]{Lebonnois2017} 
{Lebonnois}, S. \& {Schubert}, G. 2017, Nat.~Geo., 10, 473. 

\bibitem[Lejosne \& Kollmann (2020)]{lejosne2020}
Lejosne, S. \& P. Kollmann (2020), Radiation Belt Radial Diffusion at Earth and Beyond, Space Sci Rev, 216:19, doi:10.1007/s11214-020-0642-6

\bibitem[Li et~al.(2020)]{Li2020}
Li, C., et~al. 2020, Nature Astronomy, 4, 609


\bibitem[Lorenzato et~al. (2012)]{lorenzato2012}
Lorenzato, L., Sicard, A., and Bourdarie, S. (2012), A physical model for electron radiation belts of Saturn, J. Geophys. Res., 117, A08214, doi:10.1029/2012JA017560.

\bibitem[Louarn et~al.(2014)]{louarn14}
Louarn, P., \emph{et~al.}, 2014, \emph{Geophysical Research Letters}, 119, 4495.

\bibitem[Louarn et~al.(2018)]{louarn18}
Louarn, P., \emph{et~al.}, 2018, \emph{Geophysical Research Letters}, 45, 18, 9408.

\bibitem[Louarn et~al.(2017)]{louarn17}
Louarn, P., \emph{et~al.}, 2017,  \emph{Geophysical Research Letters}, 44, 10, 4439.

\bibitem[Louis et~al.(2021)]{louis21}
Louis,C.~K., et al., 2021, Journal of Geophysical Research: Space Physics, 126, e2021JA029435.






\bibitem[Mauk et~al.(2017)]{mauk17}
Mauk, B.~H., \emph{et~al.}, 2017, \emph{Nature}, 549, 7670, 66.

\bibitem[Mauk et~al.(2018)]{mauk18}
Mauk,B \emph{et~al.}, 2018, \emph{Geophysical Research Letters}, 45, 3, 1277.

\bibitem[Mauk et~al.(2020)]{mauk20}
Mauk,B.~H., 2020, \emph{et~al.}, \emph{Journal of Geophysical Research: Space Physics}, 125, 3, e2019JA027699.




\bibitem[Miyoshi et~al.(1999)]{miyoshi1999}
{Miyoshi}, Y., {Misawa}, H., {Morioka}, A., {Kondo}, T., {Koyama}, Y., \&
 {Nakajima}, J. 1999, Geophys. Res. Lett., 26, 9

\bibitem[Miyoshi et~al.(2018)]{miyoshi2018}
{Miyoshi}, Y., et~al. 2018, Earth, Planets, and Space, 70, 101




\bibitem[Morioka et~al.(2012)]{morioka12}
Morioka, A., Miyoshi, Y., Kitamura, N., Misawa, H., Tsuchiya, F., Menietti, J.,
  \& Honary, F. 2012, Journal of Geophysical Research: Space Physics, 117



\bibitem[Murakami et~al.(2016)]{murakami2016}
Murakami, G., et al. (2016), Response of Jupiter's inner magnetosphere to the solar wind derived from extreme ultraviolet monitoring of the Io plasma torus, Geophys. Res. Lett., 43, 12,308– 12,316, doi:10.1002/2016GL071675.








\bibitem[Prang\'{e} et~al.(2004)]{prange04}
Prang\'{e} et al. 2004, Nature, 432, 78.



\bibitem[Roth et~al.(2014)]{roth14}
Roth, L., Saur, J., Retherford, K.~D., Strobel, D.~F., Feldman, P.~D., McGrath,
 M.~A., \& Nimmo, F. 2014, Science, 343, 171

\bibitem[Roth et~al.(2016)]{roth16}
Roth, L., et~al. 2016, Journal of Geophysical Research: Space Physics, 121,
 2143







\bibitem[Sault et~al.(2004)]{sault2004}
{Sault}, R.~J., {Engel}, Ch., \& {de Pater}, I. 2004, Icarus, 168, 336.


\bibitem[Saur et~al.(2018)]{saur18}
Saur, J., et al., 2018, \emph{Journal of Geophysical Research: Space Physics}, 123, 11, 9560.

\bibitem[Saur et~al.(2003)]{saur03}
Saur, J., A.~Pouquet, and W.~H. Matthaeus, 2003, \emph{Geophysical research letters}, 30, 5, 1260.


 





\bibitem[Southwood \& Kivelson (1987)]{southwood1987}
Southwood, D. J., and Kivelson, M. G. (1987), Magnetospheric interchange instability, J. Geophys. Res., 92(A1), 109– 116, doi:10.1029/JA092iA01p00109.


\bibitem[Spencer et~al.(1999)]{spencer99}
Spencer, J.~R., Tamppari, L.~K., Martin, T.~Z., \& Travis, L.~D. 1999,
 Science, 284, 1514










\bibitem[Takagi et~al.(2018)]{Takagi2018} 
{Takagi}, M., et~al. 2018, J.~Geophys.~Res.~Planets, 123, 335.


 




\bibitem[Tao et~al.(2015)]{tao15}
Tao, C., et al., 2015, \emph{Journal of Geophysical Research: Space Physics}, 120, 4, 2477.



\bibitem[Thomas et~al.(2004)]{thomas04}
Thomas, N., Bagenal, F., Hill, T., \& Wilson, J. 2004, Jupiter. The planet,
 satellites and magnetosphere, 1, 561




\bibitem[Trumbo et~al.(2017)]{trumbo17}
Trumbo, S.~K., Brown, M.~E., \& Butler, B.~J. 2017, The Astronomical Journal,
 154, 148

\bibitem[Trumbo et~al.(2018)]{trumbo18}
Trumbo, S.~K., Brown, M.~E., \& Butler, B.~J. 2018, The Astronomical Journal,
 156, 161

\bibitem[Tsuchiya et~al.(2011)]{tsuchiya2011}
{Tsuchiya}, F., {Misawa}, H., {Imai}, K., \& {Morioka}, A. 2011, Journal of
 Geophysical Research (Space Physics), 116, A09202







 



\bibitem[Wu \& Lee(1979)]{wulee79}
Wu, C., \& Lee, L.,1979, The Astrophysical Journal, 230, 621


\bibitem[Yao et~al.(2021)]{yao21}
Yao,Z. \emph{et~al.}, 2021, \emph{Science Advances}, 7, 28, eabf0851.




\bibitem[Yoshioka et~al.(2014)]{Yoshioka2014}
{Yoshioka}, K., et~al. 2014, Science, 345, 1581

\bibitem[Zarka (2004)]{zarka04}
Zarka, P. 2004, Advances in Space Research, 33, 2045

\bibitem[Zarka (2007)]{zarka07}
Zarka, P. 2007, Planetary and Space Science, 55, 598









\end{thebibliography}


\end{document}